\documentclass[10pt]{article}
\usepackage[a4paper]{geometry}
\usepackage[utf8]{inputenc}
\usepackage[T1]{fontenc}
\usepackage{setspace,authblk}
\usepackage{amsmath,mathrsfs,amsfonts,amsthm,amssymb}

\usepackage[english]{babel}

\usepackage{graphicx}
 
\usepackage{hyperref}

\title{\singlespace {Profit shifting under the arm's length principle}}

\author[]{Alex A.T. Rathke\thanks{NECCT/FEA-RP/USP, University of S\~ao Paulo. Avenida dos Bandeirantes 3900, 14040-905 Ribeir\~ao Preto, SP, Brazil. E-mail: \texttt{alex.rathke@alumni.usp.br}}}

\date{\today}

\theoremstyle{plain}

\newtheorem*{theorem*}{Theorem}
\newtheorem{proposition}{Proposition}
\newtheorem*{proposition*}{Proposition}

\newtheorem*{remark*}{Remark}
\newtheorem{condition}{Condition}
\newtheorem*{condition*}{Condition}

\newtheorem*{corollary*}{Corollary}

\begin{document}

\maketitle

\begin{abstract} 

This study analyses the tax-induced profit shifting behaviour of firms and the impact of governments' anti-shifting rules. We derive a model of a firm that combines internal sales and internal debt in a full profit shifting strategy, and which is required to apply the arm's length principle and a general thin capitalisation rule. We find several cases where the firm may shift profits to low-tax countries while satisfying the usual arm's length conditions in all countries. Internal sales and internal debt may be regarded either as complementary or as substitute shifting channels, depending on how the implicit concealment costs vary after changes in all transactions. We show that the cross-effect between the shifting channels facilitates profit shifting by means of accepted transfer prices and interest rates.


\end{abstract}

\noindent\textbf{Keywords:} profit shifting, transfer pricing, debt shifting, thin capitalisation, arm's length principle.
\\
\noindent\textbf{JEL Classification:} C02; D01; K34.

\section{Introduction} \label{Introduction}

Multinational firms may reduce their global taxation by transferring taxable profits from high-tax countries to low-tax countries. One usual practice is to adjust the transfer prices and outputs on internal imports and exports, resulting in the accumulation of taxable profits on countries with lower tax rates \cite{beer2020,cobham2020,riedel2018,dharmapala2014}. Another common profit shifting channel refers to loans between related parties. A firm in a low-tax country may lend money to a related party in a high-tax country, for the financial interest incurred over this loan reduces the taxable profits disclosed by the borrower \cite{goerdt2022,delis2020,schindler2016}.

Governments focus on preventing the escape of tax revenues by imposing specific anti-shifting rules over the firms' internal transactions. The baseline approach is to require that all internal transactions be consistent with the values that would be applied in transactions between independent parties, satisfying the arm's length principle \cite{oecd2022}. A further approach specific to internal debt refers to thin capitalisation rules, which establish regulatory thresholds over the tax-deductible financial interests paid to related parties\footnote{Thin capitalisation refers to the strategy of financing the activity of the firm's affiliates by means of internal debt instead of equity capital, usually with the objective to replace non-deductible dividend payments for tax-deductible interest expenses which reduce the tax base of the borrower \cite{goerdt2022,delis2020,schindler2016}.}. 

Despite the governments' regulatory measures, literature provides substantial evidence of the profit shifting behaviour of firms. Numerous studies show that the values and the direction of internal sales and internal debt of multinational firms vary systematically with the difference in tax rates across countries, and that firms disclose excessively large profits in related parties located in low-tax countries \cite{beer2020,cobham2020,riedel2018,dharmapala2014}. Most findings are obtained from firms in the U.S. and in European countries \cite{delis2020,davies2018,nicolay2017,cristea2016,saunders2015,blouin2014,buettner2012,huizinga2008,bartelsman2003,clausing2003,bernard1990}, for these countries have historically implemented the arm's length principle at the core of their transfer pricing regulations \cite{rathke2020,marques2016,zinn2014}. The existing evidence leads to the challenge to assess whether the arm's length principle is effective. 

This paper analyses the behaviour of a multinational firm that combines both internal sales and internal debt in a full profit shifting strategy. We propose the conditions where all internal transactions satisfy the arm's length principle, while the internal interest is also subjected to a thin capitalisation rule based on the earnings stripping approach\footnote{Earnings stripping rule refers to an arbitrary limit imposed over the interest expenses paid to related parties, which is commonly based on firms' pre-tax profits obtained from independent transactions \cite{goerdt2022}.}. The multinational firm may regard the internal sales and internal debt either as complementary or substitute profit shifting channels \cite{goerdt2022,schindler2016}. We find that the firm may be successful in shifting profits to low-tax countries while satisfying overall arm's length conditions in all countries. The firm may combine both internal sales and internal debt as simultaneous shifting channels under quite usual regulatory requirements. The firm may also substitute one profit shifting channel for the other if the application of the arm's length principle produces an adverse case within the firm's maximisation problem. The consideration of the shifting channels either as complements vs. substitutes depends on how the variation on the internal transactions affects the implicit concealment costs of the firm. Overall, we show the conditions where firms may adjust all internal transactions within the regulatory bounds, so to increase the gains from profit shifting.

Analytical studies focus mostly on each shifting channel separately and show cases where abusive transfer prices indeed satisfy some of the traditional arm's length conditions, e.g. \cite{choi2018,grezik2008,gox2006,schjelderup1999,samuelson1982,kant1988,bond1980}. Studies on internal debt focus on the impact of thin capitalisation rules over the investment decisions and capital structure of the firms, e.g. \cite{moen2019,mardan2017,schindler2016}. Recent studies combine transfer pricing and internal debt to analyse the conditions under which firms substitute one profit shifting channel for the other, e.g. \cite{goerdt2022,comincioli2020,grezik2017,schindler2013}, which follows from current evidence that suggests the predominance of a substitutional effect, e.g. \cite{tran2023,delis2020,kvamme2020,nicolay2017,saunders2015}. Our paper provides new analyses on the multi-channel profit shifting by firms, including the limited effect of the arm's length principle as implemented in current tax regulations. In special, we find new conditions where firms may satisfy the arm's length conditions and thin capitalisation rules, and still shift profits to low-tax countries. Our results include cases where firms may intensify the gains from profit shifting by applying all internal imports, internal exports and internal debt as complementary shifting channels. Our generalised model and original results are the main contributions of this study.

The Section \ref{The model} next presents the full model for the firm's maximisation problem, and the analysis of the firm's behaviour under the profit shifting incentive. Section \ref{Conclusion} concludes.

\section{The model} \label{The model}

\subsection{General problem and equilibria} \label{General problem and equilibria}

Assume a multinational firm with affiliates in two different countries/markets $c = \{1,2 \} $. Affiliates promote sales within their own market $s_{c} \in \mathbb{R}_{+}$, which produce revenues $R_{c} : \mathbb{R}_{+} \rightarrow \mathbb{R}_{+}$ under total costs $C_{c} : \mathbb{R}_{+} \rightarrow \mathbb{R}_{+}$. Revenues and costs of each affiliate $c$ are independent with respect to the market sales of the other affiliate $-c$, for we have $\partial R_{c} / \partial s_{-c} = 0$, $\partial C_{c} / \partial s_{-c} = 0$. We assume the usual properties regarding revenues and costs: $R_{c}(0) = 0, \partial R_{c} / \partial s_{c} = R_{c}^{\prime} > 0$, $\partial^{2} R_{c} / \partial s_{c}^{2} = R_{c}^{\prime \prime} < 0$, $C_{c}(0) \geq 0, \partial C_{c} / \partial s_{c} = C_{c}^{\prime} > 0$, $\partial^{2} C_{c} / \partial s_{c}^{2} = C_{c}^{\prime \prime} > 0$.

Besides the market sales $s_{c}$, each affiliate $c$ may produce some additional output $q_{c} \in \mathbb{R}_{+}$ and may sell it to the other affiliate, thus charging a non-negative transfer price $p_{c} \in \mathbb{R}_{+}$. The internal outputs $q_{1}, q_{2}$ influence the total costs of both affiliates, $C_{1}, C_{2}$. The transfer prices charged by each affiliate $p_{1}, p_{2}$ may be different. The internal sales from affiliate $c$ to the other affiliate $-c$ produce taxable revenues in affiliate $c$ equal to $x_{c} = p_{c} q_{c}$, while it produces deductible purchase costs equal to $x_{c}$ on the other affiliate $-c$.

Economic activity in each affiliate is financed by equity capital and by debt with banks. All return to equity is reinvested on the firm. All debt in each affiliate is subjected to an interest expense $B_{c} \in \mathbb{R}_{+}$ based on market interest rates satisfying $B_{c}^{\prime} > 0, B_{c}^{\prime \prime} \geq 0$. The interest expense $B_{c}$ reduces the profit of the affiliate $c$ that is taking the debt. Excess financial resources may be transferred between affiliates by means of a loan agreement. Assume that the affiliate 2 lends an amount of money $b \in \mathbb{R}_{+}$ to affiliate 1 and charges a non-negative interest rate $r \in \mathbb{R}_{+}$ over this amount. The internal debt $b$ affects the cost of debt in both affiliates, $B_{1}, B_{2}$\footnote{By receiving the internal debt $b$, affiliate 1 may reduce its debt with banks, therefore the interest expense in affiliate 1 is equal to $B_{1}(-b)$, where the negative sign of $b$ within the interest function indicates that higher internal debt reduces the interest expense with banks in affiliate 1. Conversely, by lending the amount $b$ to affiliate 1, affiliate 2 may need to ask more money to banks, therefore the interest expense in affiliate 2 is equal to $B_{2}(b)$, for a higher internal debt increases the interest expense with banks in affiliate 2.}. The total internal interest paid by affiliate 1 to affiliate 2 is equal to $y = rb$. The internal interest $y$ affects the profits on both affiliates.

The total profits on each affiliate are subjected to domestic taxation at a tax rate $t_{c} \in [0,1]$ in each country. Assume that the Country 1 is the high-tax country, so we have $T = t_{1} - t_{2} > 0$. The net profit in each affiliate $\pi_{c}, c = \{1,2 \}$ is equal to

\begin{equation} \label{net profits}
\begin{array}{rl}
\pi_{1} &= (1-t_{1})[R_{1}(s_{1}) - C_{1}(s_{1} + q_{1} - q_{2}) - B_{1}(-b) + x_{1} - x_{2} - y] , \\
\\ 
\pi_{2} &= (1-t_{2})[R_{2}(s_{2}) - C_{2}(s_{2} + q_{2} - q_{1}) - B_{2}(b) - x_{1} + x_{2} + y] . \\
\end{array}
\end{equation}

The firm has incentives to choose internal transactions $x_{1}, x_{2}, y$ so to shift taxable profits from Country 1 to Country 2, therefore to increase the total net profit of the firm, $\pi_{1} + \pi_{2}$. Assume that tax authorities in Country 1 and Country 2 observe the full values incurred by the domestic affiliate $c$ as in Eq. \ref{net profits}, and assume that both countries implement anti-shifting regulations on domestic firms. The predominant approach worldwide is to require that the transfer prices $p_{1}, p_{2}$ and the internal interest rate $r$ be consistent with the values that would be applied in transactions between independent parties, following the arm's length principle \cite{oecd2022}. 

For our analysis, assume that countries independently assess the full internal transactions $x_{1}, x_{2}, y$ such that larger outflows of taxable profits are more likely to be challenged. Either Country 1 or 2 may charge taxes over alleged shifted profits, and additional tax fines may apply. Besides, the firm may spend substantial efforts to convince tax authorities that the preferred values of $p_{1}, p_{2}, r$ satisfy the arm's length principle, especially if the chosen values produce the intended shifting effect. Overall, the existence of internal transactions $x_{1}, x_{2}, y$ produce implicit costs for the firm, which are so called concealment costs \cite{riedel2018}. For any amount of internal transaction $x \in \mathbb{R}$ that reduces the taxable profits in Country $c$, define the monotone increasing functions of $x$ equal to $f,g : \mathbb{R}_{+} \rightarrow \mathbb{R}_{+}$, so $f,g$ are the concealment costs incurred by affiliates 1 and 2 respectively. The usual properties of cost functions apply: for all $\forall x \in \mathbb{R}$, $\partial f / \partial x > 0, \partial^{2} f / \partial x^{2} \geq 0, \partial g / \partial x > 0, \partial^{2} g / \partial x^{2} \geq 0$. From Eq. \ref{net profits}, our definition of concealment costs imply that we have $f = f(- x_{1}, x_{2}, y), g = g(x_{1}, - x_{2}, -y)$, where the sign of the variables within the concealment costs indicate whether the internal transactions produce an inflow (negative) or an outflow (positive) of taxable profits from Country $c$. We use the index notation for the derivatives of the concealment costs $f,g$. We have

\begin{equation} \label{df dg}
\begin{array}{l}
\dfrac{\partial f}{\partial x_{1}} = - f_{1} < 0 , \quad \dfrac{\partial f}{\partial x_{2}} = f_{2} > 0, \quad \dfrac{\partial f}{\partial y} = f_{y} > 0 , \\
\\
\dfrac{\partial g}{\partial x_{1}} = g_{1} > 0 , \quad \dfrac{\partial g}{\partial x_{2}} = - g_{2} < 0 , \quad \dfrac{\partial g}{\partial y} = - g_{y} < 0 . \\
\end{array}
\end{equation}

In addition to the arm's length principle, assume that the Country 1 may implement a thin capitalisation rule in order to prevent the debt shifting by means of excessive interest expenses \cite{oecd2022}. The two main approaches refer to the safe harbour rule and the earnings striping rule \cite{goerdt2022,mardan2017}. The safe harbour rule constraints the amount of internal debt $b$ on a predefined ratio based on the equity capital of the firm. If the internal debt exceeds the equity-based limit, the amount of internal interest proportional to this excess cannot be deduced from the taxable base of the affiliate 1. In our analysis where the firm may adjust the full internal interest $y = rb$, it is clear that the safe harbour rule is not effective since the firm may increase the internal interest $y$ by increasing the interest rate $r$ \footnote{A detailed analysis of the disadvantages of the thin capitalisation rule based on the safe harbour approach is presented in \cite{goerdt2022}.} \cite{mardan2017}.

For the second approach, the earnings striping rule constraints the whole amount of internal interest $y$ on a proportion of the firm's operational earnings before interest and taxes (EBIT) \cite{goerdt2022,mardan2017}. To simplify our analysis, assume that the earnings striping rule imposed by Country 1 is based on uncontrolled market transactions only, $s_{1}$, so that for any value of uncontrolled EBIT in affiliate 1 equal to $R_{1}(s_{1}) - C_{1}(s_{1})$, define an earnings striping function equal to $\bar{y} : \mathbb{R} \rightarrow \mathbb{R}_{+}$ which establishes a regulatory constraint on the amount of internal interest equal to $y < \bar{y}$. If the constraint is slack, $y < \bar{y}$, then the internal interest $y$ may be fully deduced from the taxable profit in affiliate 1. If the constraint is binding, $y \geq \bar{y}$, then the exceeding amount of internal interest equal to $y - \bar{y} > 0$ cannot be deduced from the tax base in Country 1 \cite{goerdt2022}. 

Assume that both Countries 1 and 2 implement their anti-shifting regulations independently, and assume countries do not cooperate with each other to solve any double-tax disputes. It means that both concealment costs $f,g$ affect the maximisation problem of the firm. Besides, the total internal interest $y$ is always taxed at full by the affiliate 2, therefore a binding constraint $y \geq \bar{y}$ produces a double taxation of the excess internal interest. From the total net profit of the firm in Eq. \ref{net profits}, the maximisation object of the firm shapes into a traditional Lagrangian equal to

\begin{equation} \label{l net profits}
\begin{array}{rl}
L &= \pi_{1} + \pi_{2} - f(-x_{1},x_{2},y) - g(x_{1}, -x_{2}, -y) - \lambda t_{1} (y - \bar{y}) , \\
\end{array}
\end{equation}

\noindent where $\lambda : \mathbb{R} \rightarrow \{0,1 \} $ is the Lagrange indicator function with value 1 if the thin capitalisation rule is binding, $y \geq \bar{y}$, and value zero otherwise. 

For convenient notation, define the difference in the net marginal costs and the net interest expenses respectively equal to $C^{\prime} = (1 - t_{1}) C^{\prime}_{1} - (1 - t_{2}) C_{2}^{\prime}, B^{\prime} = (1 - t_{1}) B_{1}^{\prime} - (1 - t_{2}) B_{2}^{\prime}$. Remark that we assume that Country 1 is the high-tax country, so we have $T = t_{1} - t_{2} > 0$. For the maximisation problem $\text{max}_{x_{1}, x_{2}, y} L$, subjected to $y < \bar{y}$, the first-order conditions imply\footnote{We have $q_{1} = x_{1} / p_{1}, q_{2} = x_{2} / p_{2}, b = y/r$ by definition, therefore the derivatives of the cost functions $C_{1}, C_{2}$ and the interest expenses $B_{1}, B_{2}$ within the net profits in Eq. \ref{net profits} are equal to

\begin{equation*}
\begin{array}{cc}
\dfrac{\partial C_{1}}{\partial x_{1}} = \dfrac{C_{1}^{\prime}}{p_{1}} , & \dfrac{\partial C_{1}}{\partial x_{2}} = - \dfrac{C_{1}^{\prime}}{p_{2}} , \\
\\
\dfrac{\partial C_{2}}{\partial x_{2}} = \dfrac{C_{2}^{\prime}}{p_{2}} , & \dfrac{\partial C_{2}}{\partial x_{1}} = - \dfrac{C_{2}^{\prime}}{p_{1}} , \\
\\
\dfrac{\partial B_{2}}{\partial b} = \dfrac{B_{2}^{\prime}}{r} , & \dfrac{\partial B_{1}}{\partial b} = - \dfrac{B_{1}^{\prime}}{r} . \\
\end{array} \\
\end{equation*}
}

\begin{equation} \label{dl full}
\begin{array}{rl}
\dfrac{\partial L}{\partial x_{1}} = L_{1} &= - T - \dfrac{C^{\prime}}{p_{1}} + f_{1} - g_{1} = 0 , \\
\\
\dfrac{\partial L}{\partial x_{2}} = L_{2} &= T + \dfrac{C^{\prime}}{p_{2}} - f_{2} + g_{2} = 0 , \\
\\
\dfrac{\partial L}{\partial y} = L_{y} &= T + \dfrac{B^{\prime}}{r} - f_{y} + g_{y} - \lambda t_{1} = 0 , \\
\\
\dfrac{\Delta L}{\Delta \lambda} = L_{\lambda} &= - t_{1} (y - \bar{y}) \geq 0 . \\
\end{array} \\
\end{equation}

The conditions regarding $L_{1}, L_{2}, L_{y}$ in Eq. \ref{dl full} refer to the existence of the equilibria points $x_{1}^{*}, x_{2}^{*}, y^{*}$ where the marginal gain from the shifting incentive $T > 0$ equates the marginal costs of profit shifting. The additional condition $L_{\lambda} \geq 0$\footnote{The Lagrange indicator $\lambda$ is a discrete variable, therefore the variation $L_{\lambda}$ is obtained by discrete difference, e.g. also called 'discrete differentiation', for we have

\begin{equation*}
\dfrac{\Delta L}{\Delta \lambda} = \dfrac{L(\lambda = 1) - L(\lambda = 0)}{(\lambda = 1) - (\lambda = 0)} = \dfrac{- t_{1} (y - \bar{y})}{1 - 0} = - t_{1} (y - \bar{y}) . \\
\end{equation*}
} qualifies the inequality constraint $y < \bar{y}$. The firm has incentives to not exceed the thin capitalisation rule so the equilibrium is equal to $y^{*} \leq \bar{y}$. 

The second-order conditions for $x_{1}^{*}, x_{2}^{*}, y^{*}, \lambda^{*}$ to be a maxima are the usual ones, i.e. the convexities $L_{11}, L_{22}, L_{yy}$ are negative and must be larger in magnitude than the cross-variations $L_{12}, L_{1y}, L_{2y}$. The full derivation of the second-order conditions for the max $x_{1}^{*}, x_{2}^{*}, y^{*}, \lambda^{*}$ is presented in the Appendix.

\subsection{Comparative analysis and the arm's length principle} \label{Structure of the comparative analysis and the arm's length principle}

Let the firm choose the max so we obtain the implicit functions $x_{1}^{*} = x_{1}(T), x_{2}^{*} = x_{2}(T), y^{*} = y(T), \lambda^{*} = \lambda(T)$. For convenience, we apply the index notation for the derivatives with respect to $T$, and we indicate the sign of the second derivatives of $L$ in a sign exponent\footnote{For example, the second derivative $\partial^{2} L / \partial x_{1}^{2} = L_{11} < 0$ is negative, so we indicate its sign by the notation $L_{11} < 0 \rightarrow L_{11}^{-}$. Further details in the Appendix.}. We apply this sign notation hereinafter for all terms with unambiguous signs.

We focus on the analysis the internal transactions $x_{1}^{*}, x_{2}^{*}, y^{*}$. The analysis of the shadow cost of thin capitalisation $\lambda^{*}$ is presented in the Appendix. Differentiate the max $x_{1}^{*}, x_{2}^{*}, y^{*}$ with respect to either tax rate $t_{1}, t_{2}$. After some manipulation, we find the solution equal to\footnote{For any two values $x,y \in \mathbb{R}$, the relation $x \sim y$ means '$x$ is sign-preserving monotonic transformation of $y$'.}

\begin{equation} \label{solution3}
\begin{array}{rl}
\dfrac{\partial x_{1}}{\partial T} + \dfrac{\partial x_{2}}{\partial T} = x_{1T} + x_{2 T} &\sim \left( \dfrac{C_{c}^{\prime}}{p_{1}} - \dfrac{C_{c}^{\prime}}{p_{2}} \right) + y_{T}^{+} (L_{1 y}^{+} + L_{2 y}^{-}) , \\
\\
\dfrac{\partial y}{\partial T} = y_{T}^{+} &= \dfrac{\bar{y} - y}{t_{1}} \geq 0 . \\
\end{array}
\end{equation}

The variable $C^{\prime}_{c} > 0$ refers to the marginal cost of the affiliate $c$, see the Lagrangian in Eq. \ref{l net profits}. The index $c$ indicates whether the profit shifting incentive $T$ varies with respect to either tax rate $t_{1}, t_{2}$. The full derivation of Eq. \ref{solution3} is presented in the Appendix. 

First of all, observe that Eq. \ref{solution3} solves for the variation on the total internal sales $x_{1 T} + x_{2 T}$ since all well-defined solutions at $x_{1}^{*}, x_{2}^{*}, y^{*}$ imply that both variations $x_{1 T}, x_{2 T}$ go on the same direction. 

\begin{proposition} \label{proposition internal sales}
Assume the general case for the maxima $x_{1}^{*}, x_{2}^{*}, y^{*}$. If the solution for the variations on the internal sales $x_{1 T}, x_{2 T}$ is well-defined, then the variation on the total internal sales $x_{1 T} + x_{2 T}$ provides the same direction information (e.g. the same sign) as each individual variation $x_{1 T}, x_{2 T}$.
\end{proposition}

The proof of Proposition \ref{proposition internal sales} is presented in the Appendix.

Eq. \ref{solution3} shows that the variation on the internal interest $y_{T}^{+} \geq 0$ is non-negative, while the variation on the total internal sales $x_{1T} + x_{2 T}$ is ambiguous. Following the analysis of \cite{schindler2016}, the internal values $x_{2}, y$ earned by affiliate 2 are regarded as substitute shifting channels with respect to each other, while each internal transaction $x_{2}, y$ is regarded as a complementary shifting channel with respect to the internal sales $x_{1}$ earned by affiliate 1\footnote{\cite{schindler2016} define concealment cost substitutability when the marginal concealment cost of one shifting channel increases after the increase of the other shifting channel. Likewise, \cite{schindler2016} define concealment cost complementarity when the marginal concealment cost of one shifting channel decreases after the increase of the other shifting channel. It follows directly from our analysis that we have

\begin{equation*}
\begin{array}{rl}
\dfrac{\partial^{2} (f + g)}{\partial x_{1} \partial x_{2}} &= - (f_{12} + g_{12}) < 0 , \\
\\
\dfrac{\partial^{2} (f + g)}{\partial x_{1} \partial y} &= - (f_{1y} + g_{1y}) < 0 , \\
\\
\dfrac{\partial^{2} (f + g)}{\partial x_{2} \partial y} &= f_{2 y} + g_{2 y} > 0 . \\
\end{array} \\
\end{equation*}
}. The relations between the internal transactions $x_{1}, x_{2}, y$ are reflected in the cross-effects\footnote{The derivation of all cross-effects is presented in Eq. \ref{ddl full} in the Appendix.} equal to $L_{12}^{+} > 0, L_{1 y}^{+} \geq 0, L_{2 y}^{-} \leq 0$.

There are two key outcomes in Eq. \ref{solution3}. First, the variation on the total internal sales $x_{1 T} + x_{2 T}$ depends directly on the relation between the transfer prices $p_{1}, p_{2}$. In this regard, notice that the internal sales $x_{1}, x_{2}$ are not constrained in the strict sense, however Countries 1 and 2 require that the transfer prices $p_{1}, p_{2}$ be consistent with the arm's length principle \cite{oecd2022}. It implies two conditions.

\begin{condition} \label{assumption1}
The arm's length principle implies the condition $C^{\prime}_{c} \leq p_{c}$. We call it 'cost-plus' condition for affiliate $c$.
\end{condition}

\begin{condition} \label{assumption2}
The arm's length principle implies the condition $p_{-c} \leq C_{c}^{\prime}$. We call it 'resale-price' condition for affiliate $c$.
\end{condition}

Condition \ref{assumption1} refers to the trivial condition that the affiliate $c$ aims to obtain a profit from sales to independent parties, therefore any arm's length transfer price $p_{c}$ would also produce a profit in that same affiliate $c$, $p_{c} - C_{c}^{\prime} \geq 0$. This is the basis of the 'cost-plus method' defined in the current OECD Transfer Pricing Guidelines \cite{oecd2022}\footnote{Condition \ref{assumption1} includes the equality $C^{\prime}_{c} = p_{c}$. This instance is also related to the 'comparable uncontrolled price' method defined in the OECD Guidelines, for the affiliate $c$ maximises its profit at the equilibrium $s_{c}^{*}$, which implies $R_{c}^{\prime} = C_{c}^{\prime} = p_{c}$.}. 

Condition \ref{assumption2} refers to the choice of the lowest cost by affiliate $c$, by comparing its own marginal cost $C_{c}^{\prime}$ and the transfer price $p_{-c}$ charged by the other affiliate $-c$. For example, if it is cheaper to produce output $q_{c}$ in Country 2 and to bring it to Country 1 at a transfer price $p_{2}$, rather than producing it in the affiliate 1, meaning that we have $p_{2} \leq C_{1}^{\prime}$, then the firm has a legitimate economic reason to do so\footnote{This condition is consistent with the arm's length principle on the point of view of the Country $c$, since the usual budgeting procedure that would be applied by affiliate $c$ with third parties leads to the same choice of the lowest cost.}. Condition \ref{assumption2} relates to the 'resale-price method' defined in the OECD Guidelines \cite{oecd2022}. To observe this, notice that the affiliate $c$ maximises its profit at the equilibrium $s_{c}^{*}$ where we have $R_{c}^{\prime} = C_{c}^{\prime}$. Condition \ref{assumption2} follows directly from this maximum, for both affiliates 1 and 2.

For the second outcome in Eq. \ref{solution3}, the effect of the relation between the internal sales $x_{1}, x_{2}$ and the internal interest $y$ is quite intuitive, see \cite{schindler2016}. Firstly, the internal sales $x_{1}, x_{2}$ are complementary shifting channels with respect to each other, $L_{12}^{+} > 0$, so it is clear that both variations $x_{1 T}, x_{2 T}$ go on the same direction, see Proposition \ref{proposition internal sales}. For the effect of the internal interest $y$, if the complementary effect dominates the solution, so we have $L_{1 y}^{+} + L_{2 y}^{-} \geq 0$, then the non-negative variation $y_{T}^{+} \geq 0$ implies an increase on the total internal sales, $x_{1 T} + x_{2 T} \geq 0$. The opposite occurs if the substitute effect dominates the solution, so we have $L_{1 y}^{+} + L_{2 y}^{-} \leq 0$, i.e. in this case, the shifting channels substitute one another, so the non-negative variation $y_{T}^{+} \geq 0$ implies a reduction on the total internal sales, $x_{1 T} + x_{2 T} \leq 0$. The effect of the variation on the internal interest is neutralised if thin capitalisation rule is binding, $y^{*} = \bar{y} \rightarrow y_{T}^{+} = 0$. 

We analyse the variations in Eq. \ref{solution3} in the next Sections.

\subsection{Analysis under a binding constraint, $y^{*} = \bar{y}$} \label{Analysis under a binding constraint}

Assume the simple case where the thin capitalisation rule is binding, so we have $y^{*} = \bar{y} \rightarrow y_{T}^{+} = 0$. Let the affiliate 2 satisfy both the 'cost-plus' condition and the 'resale-price' condition, which implies the condition $p_{1} \leq C_{2}^{\prime} \leq p_{2}$. We obtain the solution equal to

\begin{equation} \label{solution3_A12a2}
p_{1} \leq p_{2}, y_{T}^{+} = 0 \rightarrow x_{1T} + x_{2T} \geq 0 . \\
\end{equation}

Eq. \ref{solution3_A12a2} shows the case where the variation on the total internal sales $x_{1 T} + x_{2 T} \geq 0$ is positively associated with the profit shifting incentive $T$. 

\begin{proposition} \label{proposition_A12a2}
Assume the maxima under a binding constraint $x_{1}^{*}, x_{2}^{*}, y^{*} = \bar{y}$. Assume that the affiliate 2 follows the arm's length principle so to comply with both the 'cost-plus' condition and the 'resale-price' condition. Therefore, a higher (lower) profit shifting incentive $T$ induces the firm to increase (reduce) the total internal sales, $x_{1 T} + x_{2 T} \geq 0$.
\end{proposition}

Now, let the affiliate 1 satisfy both the 'cost-plus' condition and the 'resale-price' condition, which implies the condition $p_{2} \leq C_{1}^{\prime} \leq p_{1}$. We obtain the solution equal to

\begin{equation} \label{solution3_A12a1}
p_{2} \leq p_{1}, y_{T}^{+} = 0 \rightarrow x_{1T} + x_{2T} \leq 0 . \\
\end{equation}

Eq. \ref{solution3_A12a1} shows an opposite result when compared to the solution in Eq. \ref{solution3_A12a2}, for it shows the case where the variation on the total internal sales $x_{1 T} + x_{2 T} \leq 0$ is negatively associated with the profit shifting incentive $T$.

\begin{proposition} \label{proposition_A12a1}
Assume the maxima under a binding constraint $x_{1}^{*}, x_{2}^{*}, y^{*} = \bar{y}$. Assume that the affiliate 1 follows the arm's length principle so to comply with both the 'cost-plus' condition and the 'resale-price' condition. Therefore, a higher (lower) profit shifting incentive $T$ induces the firm to reduce (increase) the total internal sales, $x_{1 T} + x_{2 T} \leq 0$.
\end{proposition}

Propositions \ref{proposition_A12a2} and \ref{proposition_A12a1} show that different arm's length conditions may produce opposite effects on the total internal sales. On the one hand, the firm clearly prefers a high transfer price in affiliate 2, $p_{1} \leq p_{2}$, since our analysis assumes that the Country 1 is the high-tax country, $T = t_{1} - t_{2} > 0$. Proposition \ref{proposition_A12a2} indicates that there are combinations of the arm's length conditions for affiliates 1 and 2 which produce the preferred price inequality $p_{1} \leq p_{2}$. For example, Proposition \ref{proposition_A12a2} includes the case where both affiliates 1 and 2 satisfy the 'cost-plus' condition simultaneously, so we have $C_{1}^{\prime} \leq p_{1} \leq C_{2}^{\prime} \leq p_{2}$. Proposition \ref{proposition_A12a2} also includes the case where both affiliates 1 and 2 satisfy the 'resale-price' condition simultaneously, so we have $p_{1} \leq C_{2}^{\prime} \leq p_{2} \leq C_{1}^{\prime}$. 

If the Proposition \ref{proposition_A12a2} applies, then the 'cost-plus' condition in affiliate 2 equal to $C_{2}^{\prime} \leq p_{2}$ guarantees the accumulation of profits in the low-tax Country 2. Moreover, we have the negative variation $\partial C_{2} / \partial x_{1} < 0$, thus the 'resale-price' condition in affiliate 2 equal to $p_{1} \leq C_{2}^{\prime}$ allows the firm to combine the cost schedules from both affiliates 1 and 2 in order to promote the revenues in Country 2. Proposition \ref{proposition_A12a2} indicates which are the arm's length conditions to be satisfied by the firm so that higher internal sales $x_{1 T} + x_{2 T} \geq 0$ intensify the shifting of taxable profits to the low-tax Country 2.

On the other hand, Proposition \ref{proposition_A12a1} indicates an adverse case where the firm accumulates taxable profits in the high-tax Country 1. This adverse case occurs when the arm's length conditions imposed by Countries 1 and 2 produce the undesired price inequality $p_{2} \leq p_{1}$. If the firm cannot revert this price relation without escaping from the arm's length principle, at least the firm has incentives to shrink the inequality range. By adjusting both internal outputs $q_{1}, q_{2}$ so to reduce the marginal cost $C_{1}^{\prime}$ in affiliate 1, the firm may reduce the distance between the transfer prices $p_{1}, p_{2}$ by reducing $p_{1}$, at the same time it reduces the marginal profit $p_{1} - C_{1}^{\prime}$ earned by affiliate 1. In this case, lower internal sales $x_{1 T} + x_{2 T} \leq 0$ reduce the amount of taxable profits accumulated in the high-tax Country 1.

The impact of the internal sales $x_{1}, x_{2}$ within the total net profit of the firm reveals why the price inequality $p_{1} \leq p_{2}$ is preferred. For ease of analysis, assume zero fixed costs, $C_{1}(0) = C_{2}(0) = 0$. For the special case where the internal outputs in both affiliates are the same\footnote{In our model, the equality of internal outputs $q_{1} = q_{2}$ refers to a case of 'paper transactions', i.e. identical internal outputs $q_{1} = q_{2}$ do not affect the cost of sales in either affiliate, for we have $C_{1}(s_{1} + q_{1} - q_{2}) = C_{1}(s_{1}), C_{2}(s_{2} + q_{2} - q_{1}) = C_{2}(s_{2})$. The generalisation of Eq. \ref{l net profits q} for different internal outputs $q_{1}, q_{2}$ is straightforward.}, $q_{1} = q_{2} = q$, the net result from the internal sales $x_{1}, x_{2}$ is equal to

\begin{equation} \label{l net profits q}
L(s_{1}=0, s_{2} = 0, q) = T[ (p_{2} - p_{1})q ] - f - g . \\
\end{equation}

In Eq. \ref{l net profits q}, the firm obtains a gain from profit shifting only if the transfer price $p_{2}$ in affiliate 2 is higher than $p_{1}$, so to effectively shift taxable profits to the low-tax Country 2 while carrying implicit concealment costs $f,g$. In this case with $p_{1} < p_{2}$, Eq. \ref{solution3_A12a2} shows that the firm has incentives to increase the total internal sales, $x_{1 T} + x_{2 T} \geq 0$, see Proposition \ref{proposition_A12a2}. 


\subsection{Analysis of the general case with $y^{*} \leq \bar{y}$} \label{Analysis of the general case with}

Analyse the general constrained case with $y^{*} \leq \bar{y} \rightarrow y_{T}^{+} \geq 0$. Let the affiliate 2 satisfy both the 'cost-plus' condition and the 'resale-price' condition, which implies the condition $p_{1} \leq C_{2}^{\prime} \leq p_{2}$. The solution in Eq. \ref{solution3} is well-defined if the complementary effect dominates the solution, so we have $L_{1 y}^{+} + L_{2 y}^{-} \geq 0$. We obtain the solution equal to

\begin{equation} \label{solution3_A12a2y}
p_{1} \leq p_{2}, L_{1 y}^{+} + L_{2 y}^{-} \geq 0 \rightarrow x_{1T} + x_{2T} \geq 0 . \\
\end{equation}

Eq. \ref{solution3_A12a2y} shows a general case where the variation on the total internal sales $x_{1T} + x_{2T} \geq 0$ is positively associated with the profit shifting incentive $T$.

\begin{proposition} \label{proposition_A12a2y}
Assume the general case for the maxima $x_{1}^{*}, x_{2}^{*}, y^{*}$. Assume that the affiliate 2 follows the arm's length principle so to comply with both the 'cost-plus' condition and the 'resale-price' condition. If the complementary effect dominates the solution, so we have $L_{1 y}^{+} + L_{2 y}^{-} \geq 0$, then a higher (lower) profit shifting incentive $T$ induces the firm to increase (reduce) the total internal sales, $x_{1 T} + x_{2 T} \geq 0$.
\end{proposition}

Now, let the affiliate 1 satisfy both the 'cost-plus' condition and the 'resale-price' condition, which implies the condition $p_{2} \leq C_{1}^{\prime} \leq p_{1}$. The solution in Eq. \ref{solution3} is well-defined if the substitute effect dominates the solution, so we have $L_{1 y}^{+} + L_{2 y}^{-} \leq 0$. We obtain the solution equal to

\begin{equation} \label{solution3_A12a1y}
p_{2} \leq p_{1}, L_{1 y}^{+} + L_{2 y}^{-} \leq 0 \rightarrow x_{1T} + x_{2T} \leq 0 . \\
\end{equation}

Eq. \ref{solution3_A12a1y} shows an opposite result when compared to the solution in Eq. \ref{solution3_A12a2y}, for it shows a general case where the variation on the total internal sales $x_{1 T} + x_{2 T} \leq 0$ is negatively associated with the profit shifting incentive $T$.

\begin{proposition} \label{proposition_A12a1y}
Assume the general case for the maxima $x_{1}^{*}, x_{2}^{*}, y^{*}$. Assume that the affiliate 1 follows the arm's length principle so to comply with both the 'cost-plus' condition and the 'resale-price' condition. If the substitute effect dominates the solution, so we have $L_{1 y}^{+} + L_{2 y}^{-} \leq 0$, then a higher (lower) profit shifting incentive $T$ induces the firm to reduce (increase) the total internal sales, $x_{1 T} + x_{2 T} \leq 0$.
\end{proposition}

Propositions \ref{proposition_A12a2y} and \ref{proposition_A12a1y} present two well-defined solutions in Eq. \ref{solution3} which combine two sets of conditions, one set referring to the arm's length principle for the transfer prices $p_{1}, p_{2}$, and the other referring to the complementary vs. substitute relations between the internal transactions $x_{1}, x_{2}, y$. For the internal sales alone, the conditions are similar to the analysis in Section \ref{Analysis under a binding constraint}. The preferred price inequality $p_{1} \leq p_{2}$ produces a positive effect on the total internal sales, $1 / p_{1} - 1 / p_{2} \geq 0$, while the adverse price inequality $p_{2} \leq p_{1}$ produces the opposite effect, $1 / p_{1} - 1 / p_{2} \leq 0$. The impact of the relation between the internal interest $y$ and the internal sales $x_{1}, x_{2}$ follows the direction of the dominating cross-effect $L_{1 y}^{+}, L_{2 y}^{-}$.

One possible case is the requirement that both affiliates 1 and 2 satisfy both the 'cost-plus' condition and the 'resale-price' condition simultaneously. The only feasible solution for this special case is the full equality of transfer prices and marginal costs, $p_{1} = p_{2} = C_{c}^{\prime}$. The solution in Eq. \ref{solution3} is equal to

\begin{equation} \label{solution3_A12a12y}
p_{1} = p_{2} \rightarrow x_{1T} + x_{2T} \sim y_{T}^{+}(L_{1 y}^{+} + L_{2 y}^{-}) . \\
\end{equation}

Eq. \ref{solution3_A12a12y} shows that if the transfer prices in both affiliates 1 and 2 are the same, $p_{1} = p_{2}$, then the variation on the total internal sales $x_{1T} + x_{2T}$ is defined by the dominating cross-effect $L_{1 y}^{+}, L_{2 y}^{-}$ only. 

\begin{proposition} \label{proposition_A12a12y}
Assume the general case for the maxima $x_{1}^{*}, x_{2}^{*}, y^{*}$. Assume that both affiliates 1 and 2 follow the arm's length principle so to comply with both the 'cost-plus' condition and the 'resale-price' condition. Therefore, the variation on total internal sales $x_{1 T} + x_{2 T}$ follows the direction of the dominating cross-effect, either $L_{1 y}^{+}, L_{2 y}^{-}$. 
\end{proposition}

In the absence of internal loans, the impact of the profit shifting incentive $T$ over the internal sales $x_{1}, x_{2}$ is neutralised under equal transfer prices $p_{1} = p_{2}$, see Section \ref{Analysis under a binding constraint}. Proposition \ref{proposition_A12a12y} shows that this is no longer a neutral shifting strategy if the firm may also use the internal interest $y$ for debt shifting, as either a complement or a substitute of the internal sales $x_{1}, x_{2}$. For the constrained case with $y \leq \bar{y}$, the variation on the internal interest $y_{T}^{+} \geq 0$ is non-negative by the solution in Eq. \ref{solution3}. Assuming the price equality $p_{1} = p_{2}$, if the complementary effect dominates the solution, so we have $L_{1 y}^{+} + L_{2 y}^{-} \geq 0$, then the profit shifting channels complement each other so producing a positive effect $x_{1 T} + x_{2 T} \geq 0$. If the substitute effect dominates the solution, so we have $L_{1 y}^{+} + L_{2 y}^{-} \leq 0$, then the profit shifting channels substitute one another so producing a negative effect $x_{1 T} + x_{2 T} \leq 0$.

The variation on the internal interest $y_{T}^{+} \geq 0$ follows the profit shifting direction by the solution in Eq. \ref{solution3}. The variation on the total internal sales $x_{1 T} + x_{2 T}$ indicates if the firm is accumulating taxable profits in the low-tax Country 2, therefore obtaining gains from profit shifting. The negative variation $x_{1 T} + x_{2 T} \leq 0$ indicates the accumulation of profits in the high-tax Country 1, which is an adverse case for the firm. 


Overall, the solution for the general constrained case in Eq. \ref{solution3} provides relevant analysis of the profit shifting behaviour of the firms that are subjected to a thin capitalisation rule. We show that the internal sales $x_{1}, x_{2}$ between affiliates 1 and 2 are complementary shifting channels. We show that the firm may shift profits to the low-tax country by choosing values for the internal sales $x_{1}, x_{2}$ and for the internal interest $y$ that respectively satisfy the arm's length principle and the thin capitalisation rule. We show that the application of identical transfer prices on all internal sales $x_{1}, x_{2}$ is not a neutral shifting strategy if the firm regards the internal interest $y$ as a complementary or a substitute shifting channel. We also show the conditions for the firm to apply all internal transactions as simultaneous shifting channels, therefore producing profit shifting gains for the firm.


\subsection{Analysis of the unconstrained case} \label{The unconstrained case}

The general constrained maximisation problem in Sections \ref{Structure of the comparative analysis and the arm's length principle}-\ref{Analysis of the general case with} provides the necessary background for the analysis of the unconstrained case in this Section\footnote{In perspective, the reader may regard the unconstrained maximisation problem in this Section as the most general case for analysis, and the constrained maximisation problem as a special case. We observe that the constrained case in Sections \ref{Structure of the comparative analysis and the arm's length principle}-\ref{Analysis of the general case with} is completely well-defined using the Conditions \ref{assumption1} and \ref{assumption2} only, therefore providing a much simpler analysis which helps us to solve the fully unconstrained case in this Section. For convenience, we keep the constrained maximisation problem as the beginning kick-off analysis in this study, and we apply the outcomes in Sections \ref{Structure of the comparative analysis and the arm's length principle}-\ref{Analysis of the general case with} for the more complex case in this Section.}. Assume now that the high-tax Country 1 does not implement any thin capitalisation rule over the internal interest $y$. The maximisation problem of the firm is analogous to the general case in Section \ref{Structure of the comparative analysis and the arm's length principle}. Assume the Lagrangian in Eq. \ref{l net profits} and fix the condition equal to $\lambda = 0$, so the unconstrained maximisation object is equal to

\begin{equation} \label{l net profits un2}
L(\lambda = 0) = \pi_{1} + \pi_{2} - f(-x_{1},x_{2},y) - g(x_{1}, -x_{2}, -y) . \\
\end{equation}

We apply the index notation and the sign notation for the derivatives of $L(\lambda = 0)$ consistently throughout this Section. The full derivation of the first-order and second-order conditions for the unconstrained max $x_{1}^{*}, x_{2}^{*}, y^{*}$ is presented in the Appendix. The usual convexity conditions apply, i.e. the convexities $L_{11}^{-}, L_{22}^{-}, L_{yy}^{-}$ are negative and must dominate in magnitude the cross-variations $L_{12}^{+} > 0, L_{1 y}^{+} \geq 0, L_{2 y}^{-} \leq 0$. 

Let the firm choose the max so we obtain the implicit functions $x_{1}^{*} = x_{1}(T), x_{2}^{*} = x_{2}(T), y^{*} = y(T)$. Differentiating the max $x_{1}^{*}, x_{2}^{*}, y^{*}$ with respect to either tax rate $t_{1}, t_{2}$ and manipulating, we find the solution equal to

\begin{equation} \label{solution3 un2}
\begin{array}{rl}
x_{1T} + x_{2T} &\sim \left( \dfrac{C_{c}^{\prime}}{p_{1}} - \dfrac{C_{c}^{\prime}}{p_{2}} \right) + \phi \left( 1 - \dfrac{B_{c}^{\prime}}{r} \right) , \\
\\
y_{T} &\sim \left( 1 - \dfrac{B_{c}^{\prime}}{r} \right) + \phi \left( \dfrac{C_{c}^{\prime}}{p_{1}} - \dfrac{C_{c}^{\prime}}{p_{2}} \right) . \\
\end{array} \\
\end{equation}

The Proposition \ref{proposition internal sales} applies to the solution in Eq. \ref{solution3 un2}. The variables $C^{\prime}_{c} > 0, B^{\prime}_{c} > 0$ refer to the marginal cost and the marginal interest expense of the affiliate $c$ respectively, see Eq. \ref{l net profits un2}. The index $c \in \{1,2 \}$ indicates whether the profit shifting incentive $T$ varies with respect to either tax rate $t_{1}, t_{2}$. The full derivation of Eq. \ref{solution3 un2} is presented in the Appendix. Eq. \ref{solution3 un2} includes the new bounded parameter $\phi \in (-1, 1)$ defined as follows. 

For the regularised relation between the complementary vs. substitute effects equal to $-(L_{22}^{-} L_{1 y}^{+} + L_{11}^{-} L_{2 y}^{-} )$, the condition $-(L_{22}^{-} L_{1 y}^{+} + L_{11}^{-} L_{2 y}^{-} ) \geq 0$ indicates that the complementary effect dominates the solution, while the opposite condition $-(L_{22}^{-} L_{1 y}^{+} + L_{11}^{-} L_{2 y}^{-} ) \leq 0$ indicates that the substitute effect dominates. Hence, we define the continuous bounded parameter $\phi \in (-1,1)$ which satisfies the sign-preserving monotone relation equal to

\begin{equation} \label{parameters un1}
-( L_{22}^{-}L_{1 y}^{+} + L_{11}^{-} L_{2 y}^{-}) \sim \phi (x_{1T} + x_{2T}, y_{T}, L_{1 y}^{+}, L_{2 y}^{-}) \in (-1, 1) . \\
\end{equation}

The parameter $\phi = \phi (L_{1 y}^{+}, L_{2 y}^{-}, x_{1T} + x_{2T}, y_{T})$ is bounded within the open interval $(-1, 1)$, for a positive value $\phi \in (0, 1)$ indicates the dominance of the complementary effect, while a negative value $\phi \in (-1, 0)$ indicates the dominance of the substitute effect\footnote{The parameter $\phi$ is formally a parametric function of the variations $x_{1T} + x_{2T}, y_{T}$, for the parameter $\phi$ differs in magnitude only within the open interval $(-1,1)$ with respect to each variation $x_{1T} + x_{2T}, y_{T}$, while keeping the same sign. A null cross-effect implies $\phi = 0$.}. Details about the derivation of the parameter $\phi$ are presented in the Appendix.

Eq. \ref{solution3 un2} provides a direct analysis of the profit shifting behaviour of the firm. Each profit shifting channel $x_{1T} + x_{2T}, y_{T}$ has a main effect and a cross-effect. We refer to the main effect by the term which includes the main variables related with each profit shifting channel, in the first term at the right-hand side of each variation in Eq. \ref{solution3 un2}. Namely, the main effect for the variation on the total internal sales $x_{1T} + x_{2T}$ is equal to $C_{c}^{\prime}(1/p_{1} - 1/p_{2})$, while the main effect for the variation on the internal interest $y_{T}$ is equal to $1 - B_{c}^{\prime} / r$. We refer to the cross-effect by the term which is includes the parameter $\phi$, in the second term at the right-hand side of each variation in Eq. \ref{solution3 un2}. The main effect of one profit shifting channel is the cross-effect of the other profit shifting channel, which is intuitive. Each variable $p_{1}, p_{2}, r$ produces a larger effect as a main effect rather than as a cross-effect, due to the magnitude of the parameter $\phi \in (-1, 1)$, which is also intuitive.

The solution for the unconstrained case in Eq. \ref{solution3 un2} is completely ambiguous and depends on the relation between the transfer prices $p_{1}, p_{2}$, and on the relation between the internal interest rate $r$ and the marginal interest expense $B_{c}^{\prime} > 0$. Countries 1 and 2 require that all the transfer prices $p_{1}, p_{2}$ and the internal interest rate $r$ to be consistent with the arm's length principle. Conditions \ref{assumption1} and \ref{assumption2} apply for the transfer prices $p_{1}, p_{2}$, so we need a further condition for the internal interest rate $r$.

\begin{condition} \label{assumption3}
The arm's length principle implies the condition $B_{2}^{\prime} \leq r \leq B_{1}^{\prime}$. We call it 'cost-of-funds' condition.
\end{condition}

Condition \ref{assumption3} combines the requirements from both Countries 1 and 2. On the one hand, when the affiliate 1 chooses its funding sources, it makes sense to accept the internal debt $b$ only if it is less costly in comparison to debt with banks, which implies the condition $r \leq B_{1}^{\prime}$. On the other hand, the affiliate 2 would accept to lend the amount of money $b$ to affiliate 1 only if the financial cost produced by this internal debt is totally compensated by the affiliate 1, which implies the condition $B_{2}^{\prime} \leq r$. Condition \ref{assumption3} relates to the 'cost-of-funds method' in the current OECD Transfer Pricing Guidelines \cite{oecd2022}, for both affiliates 1 and 2.

Notice that the Condition \ref{assumption3} is not fully satisfied if we have $B_{1}^{\prime} < B_{2}^{\prime}$, e.g. the firm prefers a high interest rate satisfying the relation $B_{c}^{\prime} \leq r$, so to shift profits from Country 1 to Country 2. Notice also that the Condition \ref{assumption3} implies that the profit shifting incentive $T$ produces asymmetric effects on the solution in Eq. \ref{solution3 un2} deriving from changes in either tax rate $t_{1}, t_{2}$, i.e. changes in the tax rate $t_{c}$ in Country $c$ affects the marginal interest expense $B_{c}^{\prime}$ incurred by the affiliate $c$. This asymmetry is relevant for our model. We analyse the conditions for which the solution in Eq. \ref{solution3 un2} is well-defined.

Let the affiliate 2 satisfy both the 'cost-plus' condition and the 'resale-price' condition, and let the firm satisfy the 'cost-of-funds' condition. It implies the conditions $p_{1} \leq C_{2}^{\prime} \leq p_{2}, B_{2}^{\prime} \leq r \leq B_{1}^{\prime}$. If the profit shifting incentive $T$ varies with respect to the low tax rate, so we have $\partial T / \partial  t_{2} \rightarrow B_{c}^{\prime} = B_{2}^{\prime} \leq r$, then the solution in Eq. \ref{solution3 un2} is well-defined if the complementary effect dominates the solution, so we have $\phi \geq 0$. We obtain the solution equal to

\begin{equation} \label{solution3_A12a2un2_t2}
p_{1} \leq p_{2}, B_{2}^{\prime} \leq r, \phi \geq 0 \rightarrow \left\{
\begin{array}{l}
x_{1T} + x_{2T} \geq 0 , \\
y_{T} \geq 0 . 
\end{array} \right. \\ 
\end{equation}

Assuming the same arm's length conditions $p_{1} \leq C_{2}^{\prime} \leq p_{2}, B_{2}^{\prime} \leq r \leq B_{1}^{\prime}$, if the profit shifting incentive $T$ varies with respect to the high tax rate, so we have $\partial T / \partial  t_{1} \rightarrow B_{c}^{\prime} = B_{1}^{\prime} \geq r$, then the solution in Eq. \ref{solution3 un2} is well-defined if the substitute effect dominates the solution, so we have $\phi \leq 0$. We obtain the solution equal to

\begin{equation} \label{solution3_A12a2un2_t1}
p_{1} \leq p_{2}, r \leq B_{1}^{\prime}, \phi \leq 0 \rightarrow \left\{
\begin{array}{l}
x_{1T} + x_{2T} \geq 0 , \\
y_{T} \leq 0 . 
\end{array} \right. \\ 
\end{equation}

Eq. \ref{solution3_A12a2un2_t2} and \ref{solution3_A12a2un2_t1} show two cases where the variation on the total internal sales $x_{1T} + x_{2T} \geq 0$ is positively associated with the profit shifting incentive $T$, while the variation on the internal interest $y_{T}$ depends on how the profit shifting incentive $T$ affects the marginal interest expense $B_{c}^{\prime}$.

\begin{proposition} \label{proposition_A12a2un2_t2}
Assume the unconstrained case for the maxima $x_{1}^{*}, x_{2}^{*}, y^{*}$. Assume that the affiliate 2 follows the arm's length principle so to comply with both the 'cost-plus' condition and the 'resale-price' condition, and the firm complies with the 'cost-of-funds' condition. It implies two cases:

- If the complementary effect dominates the solution, so we have $\phi \geq 0$, then a higher (lower) profit shifting incentive $T$ following a change in the low tax rate $t_{2}$ induces the firm to increase (reduce) all internal transactions, $x_{1 T} + x_{2 T} \geq 0, y_{T} \geq 0$;

- If the substitute effect dominates the solution, so we have $\phi \leq 0$, then a higher (lower) profit shifting incentive $T$ following a change in the high tax rate $t_{1}$ induces the firm to increase (reduce) the total internal sales, $x_{1 T} + x_{2 T} \geq 0$, while it induces the firm to reduce (increase) the internal interest, $y_{T} \leq 0$.
\end{proposition}

Now, let the affiliate 1 satisfy both the 'cost-plus' condition and the 'resale-price' condition, and let the firm satisfy the 'cost-of-funds' condition. It implies the conditions $p_{2} \leq C_{1}^{\prime} \leq p_{1}, B_{2}^{\prime} \leq r \leq B_{1}^{\prime}$. If the profit shifting incentive $T$ varies with respect to the high tax rate, so we have $\partial T / \partial  t_{1} \rightarrow B_{c}^{\prime} = B_{1}^{\prime} \geq r$, then the solution in Eq. \ref{solution3 un2} is well-defined if the complementary effect dominates the solution, so we have $\phi \geq 0$. We obtain the solution equal to

\begin{equation} \label{solution3_A12a1un2_t1}
p_{2} \leq p_{1}, r \leq B_{1}^{\prime}, \phi \geq 0 \rightarrow \left\{
\begin{array}{l}
x_{1T} + x_{2T} \leq 0 , \\
y_{T} \leq 0 . 
\end{array} \right. \\ 
\end{equation}

Assuming the same arm's length conditions $p_{2} \leq C_{1}^{\prime} \leq p_{1}, B_{2}^{\prime} \leq r \leq B_{1}^{\prime}$, if the profit shifting incentive $T$ varies with respect to the low tax rate, so we have $\partial T / \partial  t_{2} \rightarrow B_{c}^{\prime} = B_{2}^{\prime} \leq r$, then the solution in Eq. \ref{solution3 un2} is well-defined if the substitute effect dominates the solution, so we have $\phi \leq 0$. We obtain the solution equal to

\begin{equation} \label{solution3_A12a1un2_t2}
p_{2} \leq p_{1}, B_{2}^{\prime} \leq r, \phi \leq 0 \rightarrow \left\{
\begin{array}{l}
x_{1T} + x_{2T} \leq 0 , \\
y_{T} \geq 0 . 
\end{array} \right. \\ 
\end{equation}

Eq. \ref{solution3_A12a1un2_t1} and \ref{solution3_A12a1un2_t2} show the opposite cases when compared to the solutions in Eq. \ref{solution3_A12a2un2_t2} and \ref{solution3_A12a2un2_t1}, for they show the unconstrained case where the variation on the total internal sales $x_{1T} + x_{2T} \leq 0$ is negatively associated with the profit shifting incentive $T$, while the variation on the internal interest $y_{T}$ depends on how the profit shifting incentive $T$ affects the marginal interest expense $B_{c}^{\prime}$.

\begin{proposition} \label{proposition_A12a1un2_t1}
Assume the unconstrained case for the maxima $x_{1}^{*}, x_{2}^{*}, y^{*}$. Assume that the affiliate 1 follows the arm's length principle so to comply with both the 'cost-plus' condition and the 'resale-price' condition, and the firm complies with the 'cost-of-funds' condition. It implies two cases:

- If the complementary effect dominates the solution, so we have $\phi \geq 0$, then a higher (lower) profit shifting incentive $T$ following a change in the high tax rate $t_{1}$ induces the firm to reduce (increase) all internal transactions, $x_{1 T} + x_{2 T} \leq 0, y_{T} \leq 0$;

- If the substitute effect dominates the solution, so we have $\phi \leq 0$, then a higher (lower) profit shifting incentive $T$ following a change in the low tax rate $t_{2}$ induces the firm to reduce (increase) the total internal sales, $x_{1 T} + x_{2 T} \leq 0$, while it induces the firm to increase (reduce) the internal interest, $y_{T} \geq 0$.
\end{proposition}

Propositions \ref{proposition_A12a2un2_t2} and \ref{proposition_A12a1un2_t1} combine the several conditions producing the four possible well-defined solutions in Eq. \ref{solution3 un2}, i.e. both variations $x_{1 T} + x_{2 T}, y_{T}$ may be either positive or negative, or each may alternate in sign in either direction. We may summarise the overall analysis in Propositions \ref{proposition_A12a2un2_t2} and \ref{proposition_A12a1un2_t1} in a more intuitive way. The firm prefers to choose internal values satisfying the conditions $p_{1} \leq p_{2}, B_{c}^{\prime} \leq r$, so to shift taxable profits to the low-tax Country 2. We observe that these preferred conditions produce a positive main effect in both profit shifting channels, respectively $ 1 / p_{1} - 1 / p_{2} \geq 0, 1 - B_{c}^{\prime} / r \geq 0$. Now, if the profit shifting channels are regarded as complements by the firm, so we have $\phi \geq 0$, then the cross-effect between the profit shifting channels is also positive. The overall effect is a positive relation between the profit shifting incentive $T$ and both profit shifting channels, $x_{1 T} + x_{2 T} \geq 0, y_{T} \geq 0$. It means that the full preferred conditions $p_{1} \leq p_{2}, B_{c}^{\prime} \leq r$ allow the firm to shift taxable profits to the low-tax Country 2 using both profit shifting channels simultaneously.

We may have the case where one of the preferred conditions is not satisfied, say either adverse case (not both) $p_{2} \leq p_{1}$ or $r \leq B_{c}^{\prime}$, for the preferred condition produces a positive main effect, while the adverse condition produces a negative main effect, either adverse case (not both) $ 1 / p_{1} - 1 / p_{2} \leq 0$ or $1 - B_{c}^{\prime} / r \leq 0$. In this case, the firm is better-off if the profit shifting channels may be regarded as substitutes to one another, which implies the parameter $\phi \leq 0$. The overall effect is a positive relation between the profit shifting incentive $T$ and the profit shifting channel related to the preferred condition, and a negative relation between $T$ and the profit shifting channel related to the adverse condition, see Eq. \ref{solution3 un2}. It means that the firm may substitute one profit shifting channel for the other, so to intensify the shifting channel related to the preferred condition while reducing the shifting channel related to the adverse condition.

If the arm's length principle produces a completely adverse case with the conditions $p_{2} \leq p_{1}, r \leq B_{c}^{\prime}$, then it implies a negative main effect in both profit shifting channels equal to $ 1 / p_{1} - 1 / p_{2} \leq 0, 1 - B_{c}^{\prime} / r \leq 0$. The full adverse conditions produce an accumulation of taxable profits in the high-tax Country 1. Here again, if the profit shifting channels are regarded as complements by the firm, so we have $\phi \geq 0$, then the cross-effect between the profit shifting channels is also negative. The overall effect is a negative relation between the profit shifting incentive $T$ and both profit shifting channels, $x_{1 T} + x_{2 T} \leq 0, y_{T} \leq 0$. It means that the full adverse conditions $p_{1} \leq p_{2}, B_{c}^{\prime} \leq r$ induce the firm to reduce the volume of all internal transactions simultaneously, so to reduce the accumulation of taxable profits in the high-tax Country 1.

At last, the intuition behind the asymmetric effect produced by the marginal interest expense $B_{c}^{\prime}$ in Eq. \ref{solution3 un2} is based on the 'cost-of-funds' condition equal to $B_{2}^{\prime} \leq r \leq B_{1}^{\prime}$, and based on the fact that changes in the tax rate $t_{c}$ influence the deductibility of the interest expense $B_{c}$ and not of $B_{-c}$. There is no asymmetry if the 'cost-of-funds' condition is not fully satisfied and we have $B_{c}^{\prime} \lessgtr r$ for all $c \in \{1,2 \}$. Notice that this asymmetry applies for the substitution between the profit shifting channels to be consistent, see Eq. \ref{solution3_A12a2un2_t1} and \ref{solution3_A12a1un2_t2}.

For the general analysis of the unconstrained case, the magnitudes of the terms in Eq. \ref{solution3 un2} play a relevant role. A null cross-effect implies $\phi = 0$, for it indicates that the internal sales $x_{1}, x_{2}$ and the internal interest $y$ are regarded by the firm as completely independent profit shifting channels. If both affiliates 1 and 2 are required to satisfy all the 'cost-plus' condition and the 'resale-price' condition simultaneously, then the only feasible solution at the equality of the transfer prices $p_{1} = p_{2}$ implies that the variation on the total internal sales is defined by the cross-effect only, equal to $\phi (1 - B_{c}^{\prime} / r)$. Likewise, if the 'cost-of-funds' condition is only feasible at the equality of the interest rates, $r = B_{c}^{\prime}$, then the variation on the internal interest is defined by the the cross-effect only, equal to $\phi C_{c}^{\prime} (1 / p_{1} - 1 / p_{2})$. This is equivalent to the Proposition \ref{proposition_A12a12y} generalised to the unconstrained case.

The variations on the internal transactions $x_{1T} + x_{2T}, y_{T}$ indicate the profit shifting by means of either shifting channel. The preferred conditions $p_{1} \leq p_{2}, B_{c}^{\prime} \leq r$ produce a positive main effect in each variation in Eq. \ref{solution3 un2}. Besides, the internal values $p_{1}, p_{2}, r$ produce a larger main effect since the parameter for the cross-effect is bounded within the interval $\phi \in (-1,1)$. If either one of the variations $x_{1T} + x_{2T}, y_{T}$ follows the sign of its cross-effect, then the other variation necessarily follows the sign of its own main effect.



The solution for the unconstrained case in Eq. \ref{solution3 un2} provides relevant analyses of the profit shifting behaviour of firms. We show that the firm may shift profits to the low-tax country by choosing values for the internal sales $x_{1}, x_{2}$ and for the internal interest $y$ which satisfy the arm's length principle. We show that the equality of transfer prices $p_{1} = p_{2}$ or equality of interest rates $r = B_{c}^{\prime}$ (not both) are not necessarily neutral shifting strategies. We also show the conditions for the firm to apply all internal transactions as simultaneous shifting channels, all in compliance with the arm's length conditions, therefore producing profit shifting gains for the firm.


\section{Conclusion} \label{Conclusion}

This study analyses the profit shifting behaviour of firms and the limited impact of the anti-shifting rules by governments. We show that firms may choose transfer prices, interest rates and internal loans which satisfy the usual arm's length conditions and thin capitalisation rules in all countries simultaneously, and which still provide the chance to conveniently shift taxable profits to low-tax countries. Preferred values for firms refer to export prices or interest rates in low-tax countries which are higher than the ones in high-tax countries. Firms obtain a gain from profit shifting by intensifying the internal transactions related to the preferred values, and reducing the internal transactions related to the adverse values.

One main result is that either full equality of transfer prices or full equality of interest rates do not eliminate the shifting incentive, due to the existence of a cross-effect between the profit shifting channels. It means that firms have incentives to adjust the internal sales even if all transfer prices are the same, or to adjust the internal interest even if all interest rates are the same. In this case, the direction of the adjustments depends on whether the firm regards the internal sales and internal debt as complementary vs. substitute shifting channels. If the firm cannot avoid a full adverse condition for transfer prices and internal interest, the best shifting strategy refers to minimising all internal transactions, therefore reducing the accumulation of taxable profits in high-tax countries.





\section*{Appendix} \label{Appendix}

\subsection*{Second-order conditions for the max $x_{1}^{*}, x_{2}^{*}, y^{*}, \lambda^{*}$}

We apply the index notation for derivatives as presented in Eq. \ref{dl full}, and we apply a compact notation to indicate the sign of the second derivatives of $L$. For example, the second derivative $\partial^{2} L / \partial x_{1}^{2} = L_{11} < 0$ is negative, so we indicate its sign by the notation $L_{11} < 0 \rightarrow L_{11}^{-}$. Positive second derivatives are indicated likewise, e.g. $\partial^{2} L / \partial x_{1} \partial x_{2} = L_{12} > 0 \rightarrow L_{12}^{+}$\footnote{In further detail, the sign exponent reflects all the usual properties of the reals. For the second derivatives equal to $L_{11}^{-} < 0, L_{12}^{+} > 0$, it implies e.g. $- L_{11}^{-} > 0, -L_{12}^{+} < 0, (L_{11}^{-})^{2} > 0$.}.

Twice differentiate the Lagrangian in Eq. \ref{l net profits} with respect to all variables of interest, $L = L(x_{1}, x_{2}, y, \lambda)$. Assume the simplified notation for the second derivative of the net costs and the second derivative of the net interest expenses respectively equal to $C^{\prime \prime} = (1 - t_{1})C^{\prime \prime}_{1} + (1 - t_{2})C^{\prime \prime}_{2} > 0, B^{\prime \prime} = (1 - t_{1})B_{1}^{\prime \prime} + (1 - t_{2})B_{2}^{\prime \prime} \geq 0$. We obtain

\begin{equation} \label{ddl full}
\begin{array}{rl}
L_{11}^{-} &= - \dfrac{C^{\prime \prime}}{p_{1}^{2}} - f_{11} - g_{11} < 0 , \\
\\
L_{22}^{-} &= - \dfrac{C^{\prime \prime}}{p_{2}^{2}} - f_{22} - g_{22} < 0 , \\
\\
L_{yy}^{-} &= - \dfrac{B^{\prime \prime}}{r^{2}} - f_{yy} - g_{yy} < 0 , \\
\\
L_{12}^{+} &= \dfrac{C^{\prime \prime}}{p_{1} p_{2}} + f_{12} + g_{12} > 0 , \\
\\
L_{1y}^{+} &= f_{1y} + g_{1y} \geq 0 , \\
\\
L_{2y}^{-} &= - f_{2y} - g_{2y} \leq 0 , \\
\\
L_{y \lambda}^{-} &= - t_{1} . \\
\\
\end{array}
\end{equation}

All lines in Eq. \ref{ddl full} derive directly from Eq. \ref{df dg}-\ref{dl full} in Section \ref{General problem and equilibria}. The second-order conditions with respect to $L_{11}^{-}, L_{22}^{-}, L_{yy}^{-}$ are directly satisfied. The outcomes for $L_{1y}^{+}, L_{2y}^{-}$ assume weak inequality since the cross-derivatives of the concealment costs $f,g$ may be equal to zero. The cross-derivative $L_{12}^{+}$ is strictly positive following the usual condition $C^{\prime \prime}_{c} > 0$. For convenience, we keep the sign notation for $L_{1y}^{+}, L_{2y}^{-}$ in spite of the weak inequalities.

The equilibrium $y^{*} \leq \bar{y}$ in Eq. \ref{dl full} in Section \ref{General problem and equilibria} may be either constrained or not, for a binding thin capitalisation rule $y^{*} = \bar{y}$ increases the marginal concealment cost of debt shifting by $\lambda t_{1}$. It means that the firm has two max solutions, namely $y^{*} < \bar{y} \rightarrow \lambda^{*} = 0, y^{*} = \bar{y} \rightarrow \lambda^{*} = 1$. 

If we have $y^{*} < \bar{y} \rightarrow \lambda^{*} = 0$, then the second-order conditions for $x_{1}^{*}, x_{2}^{*}, y^{*} < \bar{y}, \lambda^{*} = 0$ to be a maxima require that the determinant of the Hessian matrix $H$ of the second derivatives of $L$ to be negative, $|H| < 0$, and the Minors of $H$ which include the diagonal terms of $H$ to be all positive. The Hessian matrix is equal to

\begin{equation} \label{hessian}
H = \left(
\begin{array}{ccc}
L_{11}^{-} & L_{12}^{+} & L_{1y}^{+} \\
L_{12}^{+} & L_{22}^{-} & L_{2y}^{-} \\
L_{1y}^{+} & L_{2y}^{-} & L_{yy}^{-} 
\end{array} \right) . \\ 
\end{equation}

The Minor of $H$ equal to $M_{ij}$ is defined as the determinant of $H$ excluding the $i$-th row and the $j$-th column. Therefore, the second-order conditions for the max $x_{1}^{*}, x_{2}^{*}, y^{*} < \bar{y}, \lambda^{*} = 0$ are equal to

\begin{equation} \label{minors0}
\begin{array}{rl}
M_{11}^{+} &= L_{22}^{-} L_{yy}^{-} - (L_{2 y}^{-})^{2} > 0 , \\
\\
M_{22}^{+} &= L_{11}^{-} L_{yy}^{-} - (L_{1 y}^{+})^{2} > 0 , \\
\\
M_{33}^{+} &= L_{11}^{-} L_{22}^{-} - (L_{12}^{+})^{2} > 0 , \\
\\
|H| &= L_{11}^{-} L_{22}^{-} L_{yy}^{-} + 2 L_{12}^{+} L_{1 y}^{+} L_{2 y}^{-} - L_{11}^{-} (L_{2 y}^{-})^{2} - L_{22}^{-} (L_{1 y}^{+})^{2} - L_{yy}^{-} (L_{12}^{+})^{2} < 0 . \\
\end{array} \\
\end{equation}

The conditions in Eq. \ref{minors0} are the usual ones, i.e. the convexities $L_{11}^{-}, L_{22}^{-}, L_{yy}^{-}$ are negative and must dominate in magnitude the cross-variations $L_{12}^{+}, L_{1 y}^{+}, L_{2 y}^{-}$, so we have $|H| < 0$.

On the other hand, if we have $y^{*} = \bar{y} \rightarrow \lambda^{*} = 1$, then the second-order conditions for $x_{1}^{*}, x_{2}^{*}, y^{*} = \bar{y}, \lambda^{*} = 1$ to be a maxima require that the determinant of the bordered Hessian matrix $\bar{H}$ of the second derivatives of $L$ to be negative, $|\bar{H}| < 0$, and the Minors of $\bar{H}$ which include the diagonal terms of $\bar{H}$ and the cross-variation $L_{y \lambda}^{-} = -t_{1}$ to be all positive\footnote{Formally, we need to analyse the $n_{x} - n_{\lambda}$ Minors of the bordered Hessian matrix, where $n_{x}$ is the number of variables and $n_{\lambda}$ is the number of constraints. The maxima requires that the smallest Minor has the sign $(-1)^{n_{\lambda} + 1}$, and the further Minors alternate in sign, e.g. in our analysis, $n_{x} = 3, n_{\lambda} = 1$.}. The bordered Hessian matrix is equal to 

\begin{equation} \label{hessian bordered}
\bar{H} = \left(
\begin{array}{cccc}
L_{\lambda \lambda} & L_{1 \lambda} & L_{2 \lambda} & L_{y \lambda}^{-} \\
L_{1 \lambda} & L_{11}^{-} & L_{12}^{+} & L_{1y}^{+} \\
L_{2 \lambda} & L_{12}^{+} & L_{22}^{-} & L_{2y}^{-} \\
L_{y \lambda}^{-} & L_{1y}^{+} & L_{2y}^{-} & L_{yy}^{-} 
\end{array} \right) = \left( \\
\\
\begin{array}{cccc}
0 & 0 & 0 & - t_{1} \\
0 & L_{11}^{-} & L_{12}^{+} & L_{1y}^{+} \\
0 & L_{12}^{+} & L_{22}^{-} & L_{2y}^{-} \\
- t_{1} & L_{1y}^{+} & L_{2y}^{-} & L_{yy}^{-} 
\end{array} \right) . \\
\end{equation}

The Minor of $\bar{H}$ equal to $\bar{M}_{ij}$ is defined as the determinant of $\bar{H}$ excluding the $i$-th row and the $j$-th column. Therefore, the second-order conditions for the max $x_{1}^{*}, x_{2}^{*}, y^{*} = \bar{y}, \lambda^{*} = 1$ are equal to

\begin{equation} \label{minors bordered}
\begin{array}{rl}
\bar{M}_{22}^{+} &= - L_{22}^{-} (-t_{1})^{2} > 0 , \\
\\
\bar{M}_{33}^{+} &= - L_{11}^{-} (-t_{1})^{2} > 0 , \\
\\
|\bar{H}| &= - \left[ L_{11}^{-} L_{22}^{-} - (L_{12}^{+})^{2} \right] (-t_{1})^{2} = - M_{33}^{+} \cdot t_{1}^{2} < 0 . \\
\end{array} \\
\end{equation}

The conditions in Eq. \ref{minors bordered} are the same as in Eq. \ref{minors0}, i.e. the convexities $L_{11}^{-}, L_{22}^{-}, L_{yy}^{-}$ are negative and must dominate in magnitude the cross-variations $L_{12}^{+}, L_{1 y}^{+}, L_{2 y}^{-}$, so we have $|\bar{H}| < 0$.

\subsection*{Derivation of the solution in Eq. \ref{solution3}, proof of Proposition \ref{proposition internal sales}}

Assume the firm chooses the maxima according to the conditions in Eq. \ref{dl full}, so the internal values become implicit functions of the profit shifting incentive $T$, $x_{1}^{*} = x_{1}(T), x_{2}^{*} = x_{2}(T), y^{*} = y(T), \lambda^{*} = \lambda(T)$. Apply the index notation for the derivatives with respect to $T$, for we have $\partial x_{1} / \partial T = x_{1 T}, \partial x_{2} / \partial T = x_{2 T}, \partial y / \partial T = y_{T}, \partial \lambda / \partial T = \lambda_{T}$. We apply the sign notation as in Eq. \ref{ddl full} for all terms with unambiguous signs.

The profit shifting incentive $T = t_{1} - t_{2} > 0$ may vary with respect to either tax rate $t_{1}, t_{2}$. For our analysis, differentiate the maxima in Eq. \ref{dl full} first with respect to $t_{1}$. The full expressions are equal to\footnote{For any value $x(T)$ as a function of $T = t_{1} - t_{2}$, with the tax rate $t_{c}, c \in \{1,2 \}$, we have

\begin{equation*}
\dfrac{\partial x(T)}{\partial t_{c}} = \dfrac{d x}{d T} \cdot \dfrac{\partial T}{\partial t_{c}} . \\
\end{equation*}
}

\begin{equation} \label{ddl t1 max constrained}
\begin{array}{rl}
\dfrac{\partial L_{1}}{\partial t_{1}} &= -1 + \dfrac{C_{1}^{\prime}}{p_{1}} -(1 - t_{1}) \left[ \dfrac{C^{\prime \prime}_{1}}{p_{1}^{2}} x_{1T} - \dfrac{C^{\prime \prime}_{1}}{p_{1} p_{2}} x_{2T} \right] + (1 - t_{2}) \left[ - \dfrac{C^{\prime \prime}_{2}}{p_{1}^{2}} x_{1 T} + \dfrac{C^{\prime \prime}_{2}}{p_{1} p_{2}} x_{2 T} \right] \\
\\
&- f_{11} x_{1T} + f_{12} x_{2T} + f_{1 y} y_{T} -g_{11} x_{1T} + g_{12} x_{2T} + g_{1y} y_{T} = 0 , \\
\\
&L_{11}^{-} x_{1T} + L_{12}^{+} x_{2 T} + L_{1 y}^{+} y_{T} = 1 - \dfrac{C_{1}^{\prime}}{p_{1}} , \\
\\
\dfrac{\partial L_{2}}{\partial t_{1}} &= 1 - \dfrac{C_{1}^{\prime}}{p_{2}} + (1 - t_{1}) \left[ \dfrac{C_{1}^{\prime \prime}}{p_{1} p_{2}} x_{1 T} - \dfrac{C_{1}^{\prime \prime}}{p_{2}^{2}} x_{2 T} \right] - (1 - t_{2}) \left[ - \dfrac{C_{2}^{\prime \prime}}{p_{1} p_{2}} x_{1 T} + \dfrac{C_{2}^{\prime \prime}}{p_{2}^{2}} x_{2 T} \right] \\
\\
&+ f_{12} x_{1T} - f_{22} x_{2T} - f_{2 y} y_{T} + g_{12} x_{1 T} - g_{22} x_{2 T} - g_{2 y} y_{T} = 0 , \\
\\
&L_{12}^{+} x_{1 T} + L_{22}^{-} x_{2 T} + L_{2y}^{-} y_{T} = - 1 + \dfrac{C_{1}^{\prime}}{p_{2}} , \\
\\
\dfrac{\partial L_{y}}{\partial t_{1}} &= 1 - \dfrac{B_{1}^{\prime}}{r} - \lambda + (1 - t_{1}) \left[ - \dfrac{B^{\prime \prime}_{1}}{r^{2}} y_{T} \right] - (1 - t_{2}) \left[ \dfrac{B^{\prime \prime}_{2}}{r^{2}} y_{T} \right] \\
\\
&+ f_{1y} x_{1 T} - f_{2 y} x_{2 T} - f_{yy} y_{T} + g_{1 y} x_{1 T} - g_{2 y} x_{2 T} - g_{yy} y_{T} - t_{1} \lambda_{T} = 0 , \\
\\
&L_{1 y}^{+} x_{1 T} + L_{2 y}^{-} x_{2 T} + L_{yy}^{-} y_{T} - t_{1} \lambda_{T} = - 1 + \dfrac{B_{1}^{\prime}}{r} + \lambda , \\
\\
\dfrac{\partial L_{\lambda}}{\partial t_{1}} &= \bar{y} - y - t_{1} y_{T} = 0 , \\
\\
& y_{T} = \dfrac{\bar{y} - y}{t_{1}} \geq 0 . \\
\\
\end{array}
\end{equation}

Arranging the equations in Eq. \ref{ddl t1 max constrained} in standard matrix form, we have

\begin{equation} \label{comparative constrained t1}
\left(
\begin{array}{cccc}
0 & 0 & 0 & - t_{1} \\
0 & L_{11}^{-} & L_{12}^{+} & L_{1y}^{+} \\
0 & L_{12}^{+} & L_{22}^{-} & L_{2y}^{-} \\
- t_{1} & L_{1y}^{+} & L_{2y}^{-} & L_{yy}^{-} \\
\end{array} \right) \left( 
\begin{array}{c}
\lambda_{T} \\
x_{1T} \\
x_{2T} \\
y_{T} \\
\end{array} \right) = \left(
\begin{array}{c}
y - \bar{y} \\
1 - C_{1}^{\prime}/p_{1} \\
- 1 + C_{1}^{\prime}/p_{2} \\
- 1 + B_{1}^{\prime}/r + \lambda \\
\end{array} \right) . \\
\end{equation}

Now differentiate the maxima in Eq. \ref{dl full} with respect to $t_{2}$. We have

\begin{equation} \label{ddl t2 max constrained}
\begin{array}{rl}
\dfrac{\partial L_{1}}{\partial t_{2}} &= 1 - \dfrac{C_{2}^{\prime}}{p_{1}} -(1 - t_{1}) \left[ - \dfrac{C^{\prime \prime}_{1}}{p_{1}^{2}} x_{1T} + \dfrac{C^{\prime \prime}_{1}}{p_{1} p_{2}} x_{2T} \right] + (1 - t_{2}) \left[ \dfrac{C^{\prime \prime}_{2}}{p_{1}^{2}} x_{1 T} - \dfrac{C^{\prime \prime}_{2}}{p_{1} p_{2}} x_{2 T} \right] \\
\\
&+ f_{11} x_{1T} - f_{12} x_{2T} - f_{1 y} y_{T} + g_{11} x_{1T} - g_{12} x_{2T} - g_{1y} y_{T} = 0 , \\
\\
&L_{11}^{-} x_{1T} + L_{12}^{+} x_{2 T} + L_{1 y}^{+} y_{T} = 1 - \dfrac{C_{2}^{\prime}}{p_{1}} , \\
\\
\dfrac{\partial L_{2}}{\partial t_{2}} &= - 1 + \dfrac{C_{2}^{\prime}}{p_{2}} + (1 - t_{1}) \left[ - \dfrac{C_{1}^{\prime \prime}}{p_{1} p_{2}} x_{1 T} + \dfrac{C_{1}^{\prime \prime}}{p_{2}^{2}} x_{2 T} \right] - (1 - t_{2}) \left[ \dfrac{C_{2}^{\prime \prime}}{p_{1} p_{2}} x_{1 T} - \dfrac{C_{2}^{\prime \prime}}{p_{2}^{2}} x_{2 T} \right] \\
\\
&- f_{12} x_{1T} + f_{22} x_{2T} + f_{2 y} y_{T} - g_{12} x_{1 T} + g_{22} x_{2 T} + g_{2 y} y_{T} = 0 , \\
\\
&L_{12}^{+} x_{1 T} + L_{22}^{-} x_{2 T} + L_{2y}^{-} y_{T} = - 1 + \dfrac{C_{2}^{\prime}}{p_{2}} , \\
\\
\dfrac{\partial L_{y}}{\partial t_{2}} &= - 1 + \dfrac{B_{2}^{\prime}}{r} + (1 - t_{1}) \left[ \dfrac{B_{1}^{\prime \prime}}{r^{2}} y_{T} \right] - (1 - t_{2}) \left[ - \dfrac{B_{2}^{\prime \prime}}{r^{2}} y_{T} \right] \\
\\
&- f_{1y} x_{1 T} + f_{2 y} x_{2 T} + f_{yy} y_{T} - g_{1 y} x_{1 T} + g_{2 y} x_{2 T} + g_{yy} y_{T} + t_{1} \lambda_{T} = 0 , \\
\\
&L_{1 y}^{+} x_{1 T} + L_{2 y}^{-} x_{2 T} + L_{yy}^{-} y_{T} - t_{1} \lambda_{T} = - 1 + \dfrac{B_{2}^{\prime}}{r} , \\
\\
\dfrac{\partial L_{\lambda}}{\partial t_{2}} &= - t_{1} y_{T} = - t_{1}z \leq 0 , \\
\\
& y_{T} = z^{+} \geq 0 . \\
\end{array} \\
\end{equation}

Arranging the equations in Eq. \ref{ddl t2 max constrained} in standard matrix form, we have

\begin{equation} \label{comparative constrained t2}
\left(
\begin{array}{cccc}
0 & 0 & 0 & -t_{1} \\
0 & L_{11}^{-} & L_{12}^{+} & L_{1y}^{+} \\
0 & L_{12}^{+} & L_{22}^{-} & L_{2y}^{-} \\
-t_{1} & L_{1y}^{+} & L_{2y}^{-} & L_{yy}^{-} \\
\end{array} \right) \left( 
\begin{array}{c}
\lambda_{T} \\
x_{1T} \\
x_{2T} \\
y_{T} \\
\end{array} \right) = \left(
\begin{array}{c}
- t_{1} z^{+} \\
1 - C_{2}^{\prime}/p_{1} \\
- 1 + C_{2}^{\prime}/p_{2} \\
- 1 + B_{2}^{\prime}/r \\
\end{array} \right) . \\
\end{equation}

The vectors at the right-hand side of both Eq. \ref{comparative constrained t1} and Eq. \ref{comparative constrained t2} indicate different solutions with respect to variations in either $t_{1}, t_{2}$, which are reflected on the indexed marginal cost $C_{c}^{\prime}$ and the indexed marginal interest expense $B_{c}^{\prime}$. The parameter $z^{+} \geq 0$ in Eq. \ref{ddl t2 max constrained} follows from the unbounded solution for the variation on the internal interest $y_{T} \geq 0$. Moreover, variation on the high tax rate $t_{1}$ in Eq. \ref{comparative constrained t1} directly affects the shadow cost of thin capitalisation by $\lambda$, for this effect is absent in Eq \ref{comparative constrained t2}. 

Combine the solutions in Eq. \ref{comparative constrained t1} and Eq. \ref{comparative constrained t2} into one general system of equations,

\begin{equation} \label{comparative constrained}
\left(
\begin{array}{cccc}
0 & 0 & 0 & -t_{1} \\
0 & L_{11}^{-} & L_{12}^{+} & L_{1y}^{+} \\
0 & L_{12}^{+} & L_{22}^{-} & L_{2y}^{-} \\
-t_{1} & L_{1y}^{+} & L_{2y}^{-} & L_{yy}^{-} \\
\end{array} \right) \left( 
\begin{array}{c}
\lambda_{T} \\
x_{1T} \\
x_{2T} \\
y_{T} \\
\end{array} \right) = \left(
\begin{array}{c}
- t_{1} z^{+} \\
1 - C_{c}^{\prime}/p_{1} \\
- 1 + C_{c}^{\prime}/p_{2} \\
- 1 + B_{c}^{\prime}/r + \lambda_{1} \\
\end{array} \right) . \\
\end{equation}

The thin capitalisation rule imposed by Country 1 implies the equality $z^{+} = (\bar{y} - y)/t_{1} \geq 0$. The index $c \in \{1,2 \}$ regarding the variables $C_{c}^{\prime} > 0, B_{c}^{\prime} > 0$ indicate whether the profit shifting incentive varies with respect to either tax rate $t_{1}, t_{2}$. We define a double indicator equal to $\lambda_{1} : \{0,1 \} \times \{c = 1 \} \rightarrow \{0,1 \}$ with value 1 under both conditions that the Lagrangian indicator $\lambda$ has value 1 and the profit shifting incentive $T$ varies with respect to the high tax rate $t_{1}$ of Country 1, and with value zero otherwise. The $4 \times 4$ matrix at the left-hand side of Eq. \ref{comparative constrained} is the familiar bordered Hessian $\bar{H}$, see Eq. \ref{hessian bordered} in this Appendix.

Solve Eq. \ref{comparative constrained} directly for $y_{T} = z^{+}$ and substitute, for we obtain

\begin{equation} \label{comparative constrained subs2}
\begin{array}{rl}
\left(
\begin{array}{ccc}
0 & L_{11}^{-} & L_{12}^{+} \\
0 & L_{12}^{+} & L_{22}^{-} \\
-t_{1} & L_{1y}^{+} & L_{2y}^{-} \\
\end{array} \right) \left( 
\begin{array}{c}
\lambda_{T} \\
x_{1T} \\
x_{2T} \\
\end{array} \right) &= \left(
\begin{array}{c}
1 - C^{\prime}_{c} / p_{1} - L_{1y}^{+} y_{T}^{+} \\
- 1 + C^{\prime}_{c} / p_{2} - L_{2y}^{-} y_{T}^{+} \\
- 1 + B_{c}^{\prime}/r + \lambda_{1} - L_{yy}^{-} y_{T}^{+} \\
\end{array} \right) = \left(
\begin{array}{c}
\bar{\mu}_{1} \\
\bar{\mu}_{2} \\
\bar{\mu}_{r} \\
\end{array} \right) \\
\\
y_{T}^{+} &= \dfrac{\bar{y} - y}{t_{1}} \geq 0 . \\
\end{array} \\
\end{equation}

Solving Eq. \ref{comparative constrained subs2} for the full set $x_{1T}, x_{2T}, y_{T}^{+}, \lambda_{T}$, we obtain the solution for each individual variation equal to\footnote{For any two values $x,y \in \mathbb{R}$, the relation $x \sim y$ means '$x$ is a sign-preserving monotonic transformation of $y$'.}

\begin{equation} \label{solution1}
\begin{array}{rl}
x_{1T} &= \dfrac{- t_{1}}{\bar{M}_{14}^{-}} \cdot \left[ \bar{\mu}_{1} L_{22}^{-} - \bar{\mu}_{2} L_{12}^{+} \right] \sim \bar{\mu}_{1} L_{22}^{-} - \bar{\mu}_{2} L_{12}^{+} , \\
\\
x_{2T} &= \dfrac{- t_{1}}{\bar{M}_{14}^{-}} \cdot \left[ \bar{\mu}_{2} L_{11}^{-} - \bar{\mu}_{1} L_{12}^{+} \right] \sim \bar{\mu}_{2} L_{11}^{-} - \bar{\mu}_{1} L_{12}^{+} , \\
\\
y_{T}^{+} &= \dfrac{\bar{y} - y}{t_{1}} \geq 0 , \\
\\
\lambda_{T} &= \dfrac{1}{\bar{M}_{14}^{-}} \cdot \left[ \bar{\mu}_{1} M_{13}^{+} - \bar{\mu}_{2} M_{23}^{+} + \bar{\mu}_{r} M_{33}^{+} \right] \sim - \bar{\mu}_{1} M_{13}^{+} + \bar{\mu}_{2} M_{23}^{+} - \bar{\mu}_{r} M_{33}^{+} , \\
\end{array}
\end{equation}

\noindent where the Minor $\bar{M}_{14}^{-} < 0$ is strictly negative, and the Minors $M_{13}^{+}, M_{23}^{+}$ assume that the convexities $L_{11}^{-}, L_{22}^{-}$ dominate in magnitude the cross-variations $L_{12}^{+}, L_{1 y}^{+}, L_{2 y}^{-}$, for we have

\begin{equation} \label{solution1 minors}
\begin{array}{rl}
\bar{M}_{14}^{-} &= - t_{1} M_{33}^{+} < 0 , \\
\\
M_{13}^{+} &= L_{12}^{+} L_{2 y}^{-} - L_{22}^{-} L_{1 y}^{+} \geq 0 , \\
\\
M_{23}^{+} &= L_{11}^{-} L_{2 y}^{-} - L_{12}^{+} L_{1 y}^{+} \geq 0 . \\
\\
\end{array}
\end{equation}

We focus here on the analysis of the variations on the internal transactions $x_{1 T}, x_{2 T}, y_{T}^{+}$. A brief inspection of Eq. \ref{solution1} shows that the variations on the internal sales $x_{1 T}, x_{2 T}$ are well-defined only if the parameters $\bar{\mu}_{1}, \bar{\mu}_{2}$ have the same sign, for it implies that both variations $x_{1 T}, x_{2 T}$ go on the same direction. We have

\begin{equation} \label{solution1 parameters internal sales}
\begin{array}{rl}
\bar{\mu}_{1} \leq 0, \bar{\mu}_{2} \leq 0 &\rightarrow x_{1 T} \geq 0, x_{2 T} \geq 0 , \\
\\
\bar{\mu}_{1} \geq 0, \bar{\mu}_{2} \geq 0 &\rightarrow x_{1 T} \leq 0, x_{2 T} \leq 0 . \\
\\
\end{array}
\end{equation}

\noindent which proves the Proposition \ref{proposition internal sales} in Section \ref{Structure of the comparative analysis and the arm's length principle}. It means that if the variations $x_{1 T}, x_{2 T}$ are well-defined, then the variation on the total internal sales $x_{1 T} + x_{2 T}$ provides the same information as each individual variation $x_{1 T}, x_{2 T}$. The sum of the variations on the internal sales is equal to

\begin{equation} \label{total internal sales}
x_{1T} + x_{2T} \sim \bar{\mu}_{1} \left[ L_{22}^{-} - L_{12}^{+} \right] + \bar{\mu}_{2} \left[ L_{11}^{-} - L_{12}^{+} \right] . \\
\end{equation}

We have strictly negative values for the terms within the brackets, $L_{11}^{-} - L_{12}^{+} < 0, L_{22}^{-} - L_{12}^{+} < 0$. It means that there exists a negative real value $L_{\mu}^{-} < 0$ satisfying the interior condition

\begin{equation} \label{mean value1}
\begin{array}{rl}
L_{11}^{-} \leq L_{22}^{-} &\rightarrow L_{11}^{-} - L_{12}^{+} \leq L_{\mu}^{-} \leq L_{22}^{-} - L_{12}^{+} < 0 , \\
\\
L_{22}^{-} \leq L_{11}^{-} &\rightarrow L_{22}^{-} - L_{12}^{+} \leq L_{\mu}^{-} \leq L_{11}^{-} - L_{12}^{+} < 0 , \\
\\
\end{array}
\end{equation}

\noindent such that for the parameters $\bar{\mu}_{1}, \bar{\mu}_{2}$ having the same sign, the interior value $L_{\mu}^{-}$ solves the equality

\begin{equation} \label{mean value2}
\bar{\mu}_{1} \left[ L_{22}^{-} - L_{12}^{+} \right] + \bar{\mu}_{2} \left[ L_{11}^{-} - L_{12}^{+} \right] = (\bar{\mu}_{1} + \bar{\mu}_{2}) L_{\mu}^{-} . \\
\end{equation}

Substitute Eq. \ref{mean value2} into Eq. \ref{total internal sales}, so we obtain the solution in Eq. \ref{solution3} regarding the variation on the total internal sales $x_{1 T} + x_{2 T}$ equal to

\begin{equation} \label{solution1 approx1}
\begin{array}{rl}
x_{1T} + x_{2T} &\sim \bar{\mu}_{1} \left[ L_{22}^{-} - L_{12}^{+} \right] + \bar{\mu}_{2} \left[ L_{11}^{-} - L_{12}^{+} \right] \\
\\
&= (\bar{\mu}_{1} + \bar{\mu}_{2}) L_{\mu}^{-} \\
\\
&\sim - (\bar{\mu}_{1} + \bar{\mu}_{2}) = \left( \dfrac{C_{c}^{\prime}}{p_{1}} - \dfrac{C_{c}^{\prime}}{p_{2}} \right) + y_{T}^{+} (L_{1y}^{+} + L_{2 y}^{-}) , \\
\\
y_{T}^{+} &= \dfrac{\bar{y} - y}{t_{1}} \geq 0 . \\
\end{array}
\end{equation}

\subsection*{The unconstrained case, derivation of the solution in Eq. \ref{solution3 un2}}

In this Section, we analyse the maximisation problem of the firm assuming that Country 1 does not impose any thin capitalisation rule. We use the Lagrangian in Eq. \ref{l net profits} in Section \ref{General problem and equilibria} and we fix the condition equal to $\lambda = 0$. For convenience, we refer to the unconstrained maximisation object equal to $L(\lambda = 0)$ through this Section. The analysis of the unconstrained case is analogous to the general model in Section \ref{The model}, therefore we resume all derivations.

The maximisation object is equal to

\begin{equation} \label{l net profits un}
L(\lambda = 0) = \pi_{1} + \pi_{2} - f(-x_{1},x_{2},y) - g(x_{1}, -x_{2}, -y) . \\
\end{equation}

We apply the index notation and the sign notation for the derivatives of $L(\lambda = 0)$, consistent with its use in this complete study, see details in Section \ref{Structure of the comparative analysis and the arm's length principle} and in Eq. \ref{ddl full} in this Appendix.

The first-order conditions for the unconstrained maximisation problem $\text{max}_{x_{1}, x_{2}, y} L(\lambda = 0)$ are equal to

\begin{equation} \label{dl full un}
\begin{array}{rl}
\dfrac{\partial L(\lambda = 0)}{\partial x_{1}} = L_{1} &= - T - \dfrac{C^{\prime}}{p_{1}} + f_{1} - g_{1} = 0 , \\
\\
\dfrac{\partial L(\lambda = 0)}{\partial x_{2}} = L_{2} &= T + \dfrac{C^{\prime}}{p_{2}} - f_{2} + g_{2} = 0 , \\
\\
\dfrac{\partial L(\lambda = 0)}{\partial y} &= T + \dfrac{B^{\prime}}{r} - f_{y} + g_{y} = 0 . \\
\end{array} \\
\end{equation}

The second-order conditions regarding the second derivatives of $L(\lambda = 0)$ are the same as in Eq. \ref{ddl full} in this Appendix. Moreover, the equilibria $x_{1}^{*}, x_{2}^{*}, y^{*}$ is a maxima if the determinant of the Hessian matrix $H$ of the second derivatives of $L(\lambda = 0)$ is negative, $|H| < 0$, and the Minors of $H$ which include the diagonal terms of $H$ are all strictly positive. The Hessian matrix is equal to

\begin{equation} \label{hessian}
H = \left(
\begin{array}{ccc}
L_{11}^{-} & L_{12}^{+} & L_{1y}^{+} \\
L_{12}^{+} & L_{22}^{-} & L_{2y}^{-} \\
L_{1y}^{+} & L_{2y}^{-} & L_{yy}^{-} 
\end{array} \right) . \\ 
\end{equation}

The Minor of $H$ equal to $M_{ij}$ is defined as the determinant of $H$ excluding th $i$-th row and the $j$-th column. We have strictly positive diagonal Minors $M_{11}^{+}, M_{22}^{+}, M_{33}^{+}$. It implies the usual second-order conditions, i.e. the convexities $L_{11}^{-}, L_{22}^{-}, L_{yy}^{-}$ are negative and must dominate in magnitude the cross-variations $L_{12}^{+}, L_{1y}^{+}, L_{2y}^{-}$. 

Let the maxima be attained, therefore we have the implicit functions $x_{1}^{*} = x_{1}(T), x_{2}^{*} = x_{2}(T), y^{*} = y(T)$. Differentiating the maxima with respect to either $t_{1}, t_{2}$, we obtain the system of linear equations 

\begin{equation} \label{comparative}
\left(
\begin{array}{ccc}
L_{11}^{-} & L_{12}^{+} & L_{1y}^{+} \\
L_{12}^{+} & L_{22}^{-} & L_{2y}^{-} \\
L_{1y}^{+} & L_{2y}^{-} & L_{yy}^{-} \\
\end{array} \right) \left( 
\begin{array}{c}
x_{1T} \\
x_{2T} \\
y_{T} \\
\end{array} \right) = \left(
\begin{array}{c}
1 - C^{\prime}_{c} / p_{1} \\
- 1 + C^{\prime}_{c} / p_{2} \\
-1 + B^{\prime}_{c} / r \\
\end{array} \right) = \left(
\begin{array}{c}
\mu_{1} \\
\mu_{2} \\
\mu_{r} \\
\end{array} \right) . \\
\end{equation}

The index $c \in \{1,2 \}$ regarding the variables $C_{c}^{\prime} > 0, B_{c}^{\prime} > 0$ indicate whether the profit shifting incentive varies with respect to either tax rate $t_{1}, t_{2}$. The $3 \times 3$ matrix at the left-hand side of Eq. \ref{comparative} is the Hessian matrix $H$. Solving the system in Eq. \ref{comparative} for the variations $x_{1 T}, x_{2 T}, y_{T}$, we find

\begin{equation} \label{solution1 un}
\begin{array}{rl}
x_{1 T} &= \dfrac{1}{|H|} \cdot \left[ \mu_{1} M_{11}^{+} - \mu_{2} M_{12}^{-} + \mu_{r} M_{13}^{+} \right] \sim - \mu_{1} M_{11}^{+} + \mu_{2} M_{12}^{-} - \mu_{r} M_{13}^{+} , \\
\\
x_{2 T} &= \dfrac{1}{|H|} \cdot \left[ -\mu_{1} M_{12}^{-} + \mu_{2} M_{22}^{+} - \mu_{r} M_{23}^{+} \right] \sim \mu_{1} M_{12}^{-} - \mu_{2} M_{22}^{+} + \mu_{r} M_{23}^{+} , \\
\\
y_{T} &= \dfrac{1}{|H|} \cdot \left[ \mu_{1} M_{13}^{+} - \mu_{2} M_{23}^{+} + \mu_{r} M_{33}^{+} \right] \sim - \mu_{1} M_{13}^{+} + \mu_{2} M_{23}^{+} - \mu_{r} M_{33}^{+} , \\
\end{array} \\
\end{equation}

\noindent where the Minors $M_{12}^{-} \leq 0, M_{13}^{+} \geq 0, M_{23}^{+} \geq 0$ assume that the convexities $L_{11}^{-}, L_{22}^{-}, L_{yy}^{-}$ dominate in magnitude the cross-variations $L_{12}^{+}, L_{1y}^{+}, L_{2 y}^{-}$, for we have\footnote{We assume that all cross-variations are non-zero.}

\begin{equation} \label{minors}
\begin{array}{rl}
M_{12}^{-} &= L_{yy}^{-} L_{12}^{+} - L_{1 y}^{+} L_{2 y}^{-} \leq 0 , \\
\\
M_{13}^{+} &= L_{12}^{+} L_{2 y}^{-} - L_{22}^{-} L_{1y}^{+} \geq 0 , \\
\\
M_{23}^{+} &= L_{11}^{-} L_{2 y}^{-} - L_{12}^{+} L_{1y}^{+} \geq 0 . \\
\end{array} \\
\end{equation}

The signs of the parameters $\mu_{1}, \mu_{2}, \mu_{r}$ alone do not solve all individual variations $x_{1 T}, x_{2 T}, y_{T}$ simultaneously. With no additional assumptions, the unconstrained solution in Eq. \ref{solution1 un} is ambiguous.

To proceed further with the analysis of Eq. \ref{solution1 un}, we explore a similar approach as in the solution in Eq. \ref{solution3} in Section \ref{Structure of the comparative analysis and the arm's length principle}. Our initial motivation is to regard the internal sales $x_{1}, x_{2}$ as complementary shifting channels \cite{schindler2016}, i.e. the cross-effect $L_{12}^{+} > 0$ is strictly positive, see Eq. \ref{ddl full} in this Appendix. 

For convenient analysis, let all variations $x_{1 T}, x_{2 T}, y_{T}$ be regularised with respect to the diagonal Minor $M_{33}^{+}$. The diagonal Minors $M_{11}^{+}, M_{22}^{+}, M_{33}^{+}$ dominate in magnitude the adjacent Minors $M_{12}^{-}, M_{13}^{+}, M_{23}^{+}$, following the maxima conditions for the convexities $L_{11}^{-}, L_{22}^{-}, L_{yy}^{-}$. The solution in Eq. \ref{solution1 un} after regularisation with respect to the diagonal Minor $M_{33}^{+}$ is equal to

\begin{equation} \label{solution2 un}
\begin{array}{rl}
x_{1 T} &\sim - \mu_{1} \widetilde{M}_{11}^{+} + \mu_{2} \widetilde{M}_{12}^{-} - \mu_{r} \widetilde{M}_{13}^{+} , \\
\\
x_{2 T} &\sim \mu_{1} \widetilde{M}_{12}^{-} - \mu_{2} \widetilde{M}_{22}^{+} + \mu_{r} \widetilde{M}_{23}^{+} , \\
\\
y_{T} &\sim - \mu_{1} \widetilde{M}_{13}^{+} + \mu_{2} \widetilde{M}_{23}^{+} - \mu_{r} , \\
\end{array} \\
\end{equation}

\noindent where the adjacent Minors become bounded after regularisation, for we have $\widetilde{M}_{12}^{-} = M_{12}^{-} / M_{33}^{+} \in (-1, 0], \widetilde{M}_{13}^{+} = M_{13}^{+} / M_{33}^{+} \in [0, 1), \widetilde{M}_{13}^{+} = M_{23}^{+} / M_{33}^{+} \in [0, 1)$. 

The sum of the variations on the internal sales is equal to

\begin{equation} \label{total internal sales un}
x_{1T} + x_{2T} \sim \mu_{1} \left[ \widetilde{M}_{12}^{-} - \widetilde{M}_{11}^{+} \right] + \mu_{2} \left[ \widetilde{M}_{12}^{-} - \widetilde{M}_{22}^{+} \right] - \mu_{r} \left[ \widetilde{M}_{13}^{+} - \widetilde{M}_{23}^{+} \right] . \\
\end{equation}

The terms within the first two brackets at the right-hand side of Eq. \ref{total internal sales un} are strictly negative and non-dominated, $\widetilde{M}_{12}^{-} - \widetilde{M}_{11}^{+} < 0, \widetilde{M}_{12}^{-} - \widetilde{M}_{22}^{+} < 0$. If we assume that the parameters $\mu_{1}, \mu_{2}$ have the same sign\footnote{The sign of each parameter $\mu_{1}, \mu_{2}$ depends on the combination of the 'cost-plus' condition and the 'resale-price' condition for affiliates 1 and 2, see Conditions \ref{assumption1} and \ref{assumption2} in Section \ref{Structure of the comparative analysis and the arm's length principle}.}, then there exists a negative real value $M_{\mu}^{-} < 0$ satisfying the interior condition

\begin{equation} \label{mean value3}
\begin{array}{rl}
\widetilde{M}_{11}^{+} \leq \widetilde{M}_{22}^{+} &\rightarrow \widetilde{M}_{12}^{-} - \widetilde{M}_{22}^{+} \leq M_{\mu}^{-} \leq \widetilde{M}_{12}^{-} - \widetilde{M}_{11}^{+} < 0 , \\
\\
\widetilde{M}_{22}^{+} \leq \widetilde{M}_{11}^{+} &\rightarrow \widetilde{M}_{12}^{-} - \widetilde{M}_{11}^{+} \leq M_{\mu}^{-} \leq \widetilde{M}_{12}^{-} - \widetilde{M}_{22}^{+} < 0 , \\
\\
\end{array}
\end{equation}

\noindent such that the interior value $M_{\mu}^{-}$ solves the equality

\begin{equation} \label{mean value4}
\mu_{1} \left[ \widetilde{M}_{12}^{-} - \widetilde{M}_{11}^{+} \right] + \mu_{2} \left[ \widetilde{M}_{12}^{-} - \widetilde{M}_{22}^{+} \right] = (\mu_{1} + \mu_{2}) M_{\mu}^{-} . \\
\end{equation}

The solution for the interior value $M_{\mu}^{-}$ in Eq. \ref{mean value3} also dominates in magnitude the regularised adjacent Minors $\widetilde{M}_{13}^{+}, \widetilde{M}_{23}^{+} \in [0, 1)$. In special, notice that the adjacent Minors $\widetilde{M}_{13}^{+}, \widetilde{M}_{23}^{+}$ carry information about the complementary vs. substitute effects $L_{1y}^{+}, L_{2 y}^{-}$ respectively, so the difference $\widetilde{M}_{13}^{+} - \widetilde{M}_{23}^{+}$ indicates which effect dominates the solution in Eq. \ref{solution1 un}. We have

\begin{equation} \label{adjacent minors4}
\begin{array}{rl}
\widetilde{M}_{13}^{+} - \widetilde{M}_{23}^{+} &= \dfrac{L_{12}^{+} (L_{1y}^{+} + L_{2 y}^{-}) - L_{22}^{-}L_{1 y}^{+} - L_{11}^{-} L_{2 y}^{-} }{M_{33}^{+} } \\
\\
&\approx - \dfrac{ L_{22}^{-}L_{1 y}^{+} + L_{11}^{-} L_{2 y}^{-} }{M_{33}^{+}} \\
\\
&\sim - (L_{22}^{-}L_{1 y}^{+} + L_{11}^{-} L_{2 y}^{-} ) , \\
\\
\end{array}
\end{equation}

\noindent so we define the bounded parameter $\phi_{x} (L_{1 y}^{+},  L_{2 y}^{-}) = \phi_{x} \in (-1,1)$ equal to

\begin{equation} \label{adjacent minors2}
\phi_{x} (L_{1 y}^{+}, L_{2 y}^{-}) = - \dfrac{\widetilde{M}_{13}^{+} - \widetilde{M}_{23}^{+}}{M_{\mu}^{-}} \sim -( L_{22}^{-}L_{1 y}^{+} + L_{11}^{-} L_{2 y}^{-} ) . \\
\end{equation}

The condition $\phi_{x} \geq 0$ indicates that the complementary effect $L_{1 y}^{+}$ dominates the solution, while the condition $\phi_{x} \leq 0$ indicates that the substitute effect $L_{2 y}^{-}$ dominates.

The bounded condition for the regularised adjacent Minors $\widetilde{M}_{13}^{+}, \widetilde{M}_{23}^{+} \in [0, 1)$ also implies that there exists a bounded real value $\phi_{y} \in (-1, 1)$ which solves the equality

\begin{equation} \label{mean value5}
\mu_{1} \widetilde{M}_{13}^{+} - \mu_{2} \widetilde{M}_{23}^{+} = (\mu_{1} + \mu_{2}) \phi_{y} . \\
\end{equation}

If the monotone condition $-L_{22}^{-}L_{1 y}^{+} - L_{11}^{-} L_{2 y}^{-} \sim \mu_{1} - \mu_{2}$ applies, then the new bounded parameter $\phi_{y} \in (-1, 1)$ also indicates which effect $L_{1y}^{+}, L_{2 y}^{-}$ dominates the solution in Eq. \ref{solution1 un}, for it satisfies the monotone relation equal to

\begin{equation} \label{adjacent minors5}
\phi_{x} (L_{1 y}^{+}, L_{2 y}^{-}) \sim \phi_{y} . \\
\end{equation}

For convenient notation, define the parametric function equal to $\phi : \mathbb{R} \rightarrow \{ \phi_{x}, \phi_{y} \}$ which is equal to $\phi = \phi_{x}$ if we refer to the variation on the total internal sales $x_{1T} + x_{2T}$, and it is equal to $\phi = \phi_{y}$ if we refer to the variation on the internal interest $y_{T}$. Finally substituting the parameter $\phi$ into Eq. \ref{total internal sales un} and manipulating, the complete solution in Eq. \ref{solution3 un2} regarding the variation on the total internal sales $x_{1 T} + x_{2 T}$ is equal to

\begin{equation} \label{solution3 un}
\begin{array}{rl}
x_{1T} + x_{2T} &\sim \mu_{1} \left[ \widetilde{M}_{12}^{-} - \widetilde{M}_{11}^{+} \right] + \mu_{2} \left[ \widetilde{M}_{12}^{-} - \widetilde{M}_{22}^{+} \right] - \mu_{r} \left[ \widetilde{M}_{13}^{+} - \widetilde{M}_{23}^{+} \right] \\
\\
&= (\mu_{1} + \mu_{2}) M_{\mu}^{-} - \mu_{r} \left[ \widetilde{M}_{13}^{+} - \widetilde{M}_{23}^{+} \right] \\
\\
&\sim - (\mu_{1} + \mu_{2}) - \phi \mu_{r} = \left( \dfrac{C_{c}^{\prime}}{p_{1}} - \dfrac{C_{c}^{\prime}}{p_{2}} \right) + \phi \left( 1 - \dfrac{B_{c}^{\prime}}{r} \right) , \\
\\
y_{T} &\sim - \mu_{1} \widetilde{M}_{13}^{+} + \mu_{2} \widetilde{M}_{23}^{+} - \mu_{r} \\
\\
&= - (\mu_{1} + \mu_{2}) \phi - \mu_{r} = \left( 1 - \dfrac{B_{c}^{\prime}}{r} \right) + \phi \left( \dfrac{C_{c}^{\prime}}{p_{1}} - \dfrac{C_{c}^{\prime}}{p_{2}} \right) , \\
\end{array} \\
\end{equation}

\noindent where the key condition for our analysis is the sign-preserving monotonic relation $-( L_{22}^{-}L_{1 y}^{+} + L_{11}^{-} L_{2 y}^{-}) \sim \phi \in (-1, 1)$.

\subsection*{Variation on the shadow cost of thin capitalisation $\lambda_{T}$}

Recall the variation on the shadow cost of thin capitalisation $\lambda_{T}$ derived in Eq. \ref{solution1} in this Appendix equal to

\begin{equation} \label{lambda t}
\begin{array}{rl}
\lambda_{T} &\sim - \bar{\mu}_{1} M_{13}^{+} + \bar{\mu}_{2} M_{23}^{+} - \bar{\mu}_{r} M_{33}^{+} \\
\\
&\sim - \bar{\mu}_{1} \widetilde{M}_{13}^{+} + \bar{\mu}_{2} \widetilde{M}_{23}^{+} - \bar{\mu}_{r} \\
\\
& = \left( \dfrac{C_{c}^{\prime}}{p_{1}} - 1 + L_{1y}^{+} y_{T}^{+} \right) \widetilde{M}_{13}^{+} - \left( 1 - \dfrac{C_{c}^{\prime}}{p_{2}} + L_{2y}^{-} y_{T}^{+} \right) \widetilde{M}_{23}^{+} \\
\\
&+ \left( 1 - \lambda_{1} - \dfrac{B_{c}^{\prime}}{r} + L_{yy}^{-} y_{T}^{+} \right) \\
\end{array} \\
\end{equation}

\noindent where the Minors $M_{13}^{+}, M_{23}^{+}, M_{33}^{+}$ assume that the convexities $L_{11}^{-}, L_{22}^{-}, L_{yy}^{-}$ dominate in magnitude the cross-variations $L_{12}^{+}, L_{1 y}^{+}, L_{2 y}^{-}$. The regularised adjacent Minors are bounded so we have $\widetilde{M}_{13}^{+} = M_{13}^{+} / M_{33}^{+}, \widetilde{M}_{23}^{+} = M_{23}^{+} / M_{33}^{+} \in [0,1)$. 

Analogous to the unconstrained case in Eq. \ref{adjacent minors4} in this Appendix, the regularised adjacent Minors $\widetilde{M}_{13}^{+}, \widetilde{M}_{23}^{+}$ carry information about the complementary vs. substitute effects $L_{1y}^{+}, L_{2 y}^{-}$ respectively, so the difference $\widetilde{M}_{13}^{+} - \widetilde{M}_{23}^{+}$ indicates which effect dominates the solution in Eq. \ref{solution3} in Section \ref{Structure of the comparative analysis and the arm's length principle}. 

If the monotone condition $-L_{22}^{-}L_{1 y}^{+} - L_{11}^{-} L_{2 y}^{-} \sim \bar{\mu}_{1} - \bar{\mu}_{2}$ applies, then we may define a bounded parameter $\phi_{\lambda}(L_{1y}^{+}, L_{2y}^{-}) = \phi_{\lambda} \in (-1,1)$ equal to 

\begin{equation} \label{adjacent minors6}
\phi_{\lambda} (L_{1 y}^{+}, L_{2 y}^{-}) = \widetilde{M}_{13}^{+} - \widetilde{M}_{23}^{+} \sim -( L_{22}^{-}L_{1 y}^{+} + L_{11}^{-} L_{2 y}^{-} ) , \\
\end{equation}

\noindent so the parameter $\phi_{\lambda}$ solves the equality

\begin{equation} \label{mean value6}
\bar{\mu}_{1} \widetilde{M}_{13}^{+} - \bar{\mu}_{2} \widetilde{M}_{23}^{+} = (\bar{\mu}_{1} + \bar{\mu}_{2}) \phi_{\lambda} . \\
\end{equation}

The condition $\phi_{\lambda} \geq 0$ indicates that the complementary effect $L_{1 y}^{+}$ dominates the solution, while the condition $\phi_{\lambda} \leq 0$ indicates that the substitute effect $L_{2 y}^{-}$ dominates. 

Substitute the bounded parameter $\phi_{\lambda} \in (-1,1)$ into Eq. \ref{lambda t} and manipulate, for we have

\begin{equation} \label{lambda t2}
\begin{array}{rl}
\lambda_{T} &\sim - \bar{\mu}_{1} \widetilde{M}_{13}^{+} + \bar{\mu}_{2} \widetilde{M}_{23}^{+} - \bar{\mu}_{r} = - (\bar{\mu}_{1} + \bar{\mu}_{2}) \phi_{\lambda} - \bar{\mu}_{r} \\
\\
&= \phi_{\lambda} \left( \dfrac{C_{c}^{\prime}}{p_{1}} - \dfrac{C_{c}^{\prime}}{p_{2}} \right) + \left( 1 - \lambda_{1} - \dfrac{B_{c}^{\prime}}{r} \right) + y_{T}^{+} \left[ \phi_{\lambda}(L_{1y}^{+} + L_{2y}^{-}) + L_{yy}^{-} \right] . \\
\end{array} \\
\end{equation}

The last term at the right-hand side of Eq. \ref{lambda t2} is dominated by the convexity $L_{yy}^{-}$, so it is always non-positive. 

Analyse first the variation under a binding constraint, $y^{*} = \bar{y}$. It implies the conditions $\lambda = 1, y_{T}^{+} = 0$, for we have

\begin{equation} \label{lambda t2b}
\lambda_{T}(y^{*} = \bar{y}) \sim \phi_{\lambda} \left( \dfrac{C_{c}^{\prime}}{p_{1}} - \dfrac{C_{c}^{\prime}}{p_{2}} \right) - \dfrac{B_{c}^{\prime}}{r} \leq 0. \\
\end{equation}

Remark that the Lagrange indicator function equal to $\lambda : \mathbb{R} \rightarrow \{0,1 \}$ has value 1 if the thin capitalisation rule is binding, $y \geq \bar{y}$, and value zero otherwise. The binding constraint at the max $y^{*} = \bar{y}$ triggers the Lagrange indicator so it reaches the upper bound of its range, $\lambda^{*} = 1$, therefore the variation on the shadow cost of thin capitalisation is non-positive by definition, $\lambda_{T}(y^{*} = \bar{y}) \leq 0$. In this case, it is only feasible for the max internal interest $y^{*} = \bar{y}$ to become slack or to be unchanged.

The solution under a binding constraint in Eq. \ref{lambda t2b} is well-defined if we have either combination of conditions $p_{1} \leq p_{2}, \phi_{\lambda} \leq 0$, or $p_{2} \leq p_{1}, \phi_{\lambda} \geq 0$. The condition regarding the preferred price relation $p_{1} \leq p_{2}$ produces a positive variation on the total internal sales $x_{1T} + x_{2T} \geq 0$, see Proposition \ref{proposition_A12a2} in Section \ref{Analysis under a binding constraint}. In this case, the solution in Eq. \ref{lambda t2b} becomes non-positive well-defined if the substitute effect dominates, so we have $\phi_{\lambda} \leq 0$. On the other hand, the condition regarding the adverse price relation $p_{2} \leq p_{1}$ produces a negative variation on the total internal sales $x_{1T} + x_{2T} \leq 0$, see Proposition \ref{proposition_A12a1} in Section \ref{Analysis under a binding constraint}. In this other case, the solution in Eq. \ref{lambda t2b} becomes non-positive well-defined if the complementary effect dominates, so we have $\phi_{\lambda} \geq 0$. 

Now assume that the constraint is slack, $y^{*} < \bar{y}$, which implies the conditions $\lambda = 0, y_{T}^{+} > 0$, so we have

\begin{equation} \label{lambda t2s}
\lambda_{T} (y^{*} < \bar{y}) \sim \phi_{\lambda} \left( \dfrac{C_{c}^{\prime}}{p_{1}} - \dfrac{C_{c}^{\prime}}{p_{2}} \right) + \left( 1 - \dfrac{B_{c}^{\prime}}{r} \right) + y_{T}^{+} \left[ \phi_{\lambda}(L_{1y}^{+} + L_{2y}^{-}) + L_{yy}^{-} \right] \geq 0 . \\
\end{equation}

The slack constraint at the max $y^{*} < \bar{y}$ implies that the Lagrange indicator is at the lower bound of it range, $\lambda = 0$, therefore the variation on the shadow cost of thin capitalisation is non-negative by definition, $\lambda_{T}(y^{*} < \bar{y}) \geq 0$. The max internal interest may vary in either direction within the thin capitalisation rule $y^{*} \leq \bar{y}$.

The solution under a slack constraint in Eq. \ref{lambda t2s} depends on further conditions regarding the transfer prices $p_{1}, p_{2}$, the internal interest rate $r$, and the parameter $\phi_{\lambda}$. Overall, the preferred conditions $p_{1} \leq p_{2}, B_{c}^{\prime} \leq r$ are associated with a dominating complementary effect, so we have $\phi_{\lambda} \geq 0$, for these conditions produce an increase in the variation on the shadow cost of thin capitalisation $\lambda_{T}$. The adverse condition $p_{2} \leq p_{1}$ is associated with a dominating substitute effect, so we have $\phi_{\lambda} \leq 0$, which also produces an increase in $\lambda_{T}$. The adverse condition $r \leq B_{c}^{\prime}$ produces a reduction in the variation $\lambda_{T}$ in Eq. \ref{lambda t2s}.

\end{document}